# Conditional Values: The Contingent Foundations of Attitudes toward Artificial Intelligence

Sam Flanders and Melati Nungsari
Asia School of Business, Kuala Lumpur, Malaysia
Correspondence: samuel.flanders@asb.edu.my

## Abstract

How do broader values shape whether people judge artificial intelligence (AI) as good or bad for humanity? This article analyzes provisional World Values Survey Wave 8 data from 40,960 respondents in 24 country samples using pooled and national prediction models, outcome-independent respondent profiles, and groups of countries. The analysis reveals a cross-country anchor: general optimism about science and technology is the strongest pooled predictor of favorable AI evaluation and recurs among the leading predictors in every national sample. Expert authority, education and information access, selected forms of trust, openness, and demographic position provide nationally distinctive patterns. Eleven value profiles formed without AI attitudes span 49.1% to 62.2% favorable evaluation. Most evaluative variation occurs within these profiles, making them useful contexts for identifying how the same values distinguish favorable from unfavorable evaluators in different social configurations. Institutional legitimacy, structural advantage, information connectedness, personal choice, redistribution, and openness differentiate evaluations in particular profiles. The same axis can relate differently to AI evaluation across profiles: secular reasoning has a significant positive association among secular moderate pluralists but a suggestive negative association among institution-trusting civic joiners. Together, the findings show that public receptivity to AI as technological change has a common predictive anchor but develops within plural configurations of resources, institutions, authority, openness, and social position. This distinction helps anticipate where generalized technological optimism is likely to travel and where AI deployment and governance will encounter context-specific sources of legitimacy or concern.

## 1. Introduction

Public attitudes toward artificial intelligence (AI) encompass judgments about both a technology and its place in society. AI can represent scientific progress, administrative capacity, surveillance, labor displacement, expert authority, concentrated power, or an alternative to failing institutions. Asking whether AI is good or bad for humanity therefore elicits a societal evaluation of technological change and the social order expected to govern it. Such evaluations are not adoption behavior, but they form part of the social environment in which AI is introduced, contested, and governed.

Most technology-acceptance research examines personal use, behavioral intention, or willingness to delegate a particular task (Kelly, Kaye, and Oviedo-Trespalacios 2023; Koenig 2025). Generalized AI evaluations are broader and often ambivalent: perceived benefits and threats can coexist, and confidence in one institution need not imply confidence in another (Schepman and Rodway 2020, 2023). Comparative work shows that science orientation, education, trust, values, and national setting matter, but it remains unclear whether these correlates form a coherent pro-AI worldview or combine differently across social contexts (O'Shaughnessy et al. 2023; Rice and Wu 2025; Wang 2025). This is also a problem for anticipating technological change: national averages can indicate broad receptivity while concealing the value configurations through which a technology acquires social legitimacy.

This article treats a broad AI evaluation as a judgment about **governed modernity**. The phrase is a sensitizing concept, not a measured scale. It directs attention to how technological change is connected to scientific progress, institutional competence, distribution, authority, information, and human control. It also motivates a distinction central to the analysis: values that define a social context need not be the same values that distinguish favorable from unfavorable AI evaluators within it.

Three questions follow. First, which non-AI characteristics provide recurrent predictive anchors across the 24 national samples, and how do those anchors combine with nationally distinctive predictors? Second, what recurring value profiles appear when AI attitudes are excluded from their construction, and which values distinguish favorable from unfavorable evaluators within each profile? Third, what broader country contexts emerge from national profile distributions and mean value positions, and how do value-evaluation associations vary across them?

A layered design answers these questions at complementary levels. Pooled prediction identifies cross-country regularities among respondents from the observed national settings. Separately fitted national models reveal which features recur and which form distinctive national combinations, while leave-one-country-out models map the reach of the common predictive structure. Respondent profiles are formed from non-AI values and social position, excluding AI attitudes and country-linked proxies; country contexts are then formed from profile shares and mean value positions. Once these outcome-independent contexts are fixed, joint models estimate which thematic dimensions distinguish AI evaluations within them.

Four results organize the article. First, general science-and-technology optimism is the clearest pooled and national predictor of favorable AI evaluation. General science optimism and AI evaluation are related but distinct: AI raises distinctive questions about autonomous decision making, opacity, surveillance, labor, and delegated authority. Second, expert decision making, education and information access, openness, demographic position, and source-specific forms of trust define a set of concrete secondary regularities. Their different national combinations show how a common science anchor is embedded in varied institutional and social settings. Third, the eleven outcome-independent respondent profiles differ by 13.1 percentage points in favorable evaluation and expose substantial within-profile differentiation. Eight thematic dimensions vary across profiles in the selected partition; science optimism is the most portable of these relationships, recurring across every corrected partition. Fourth, three country contexts reveal substantively different routes through governed modernity. The religious-traditional context reaches the highest favorable mean without high science optimism, while the secular-emancipative context shows the strongest relationship between science optimism and AI evaluation. Personal-choice liberalism also changes its association across these settings.

The article contributes a comparative map of **common anchors in plural configurations**. It identifies the clearest cross-country correlates of favorable AI evaluation, shows how their importance is recombined within recurring value profiles, and demonstrates that countries can arrive at favorable evaluations through different configurations of progress, resources, institutions, authority, and openness. For research on technological change, the analysis separates signals that travel across national settings from value relationships that remain context-dependent. This approach connects concrete predictors that readers can recognize to a broader account of how societies evaluate AI as a project of governed modernity.

## 2. Literature Review

### 2.1 From technology acceptance to societal evaluation and sociotechnical imaginaries

Much of the research on AI attitudes is still organized by frameworks developed for technology acceptance. Kelly, Kaye, and Oviedo-Trespalacios (2023), in a systematic review of user acceptance studies, show how strongly the field has been shaped by models emphasizing perceived usefulness, effort expectancy, trust, and behavioral intention. That literature is valuable, but it focuses primarily on whether people will use AI-enabled products or services. Koenig (2025) argues that this broad "acceptance" literature often conflates three analytically distinct questions: user acceptance, delegation acceptance, and societal adoption acceptance. The first concerns whether individuals will personally use AI systems; the second concerns willingness to hand decisions to AI; the third concerns whether AI is judged socially desirable at a wider collective level.

The WVS item analyzed here is closest to this third register. Respondents are not asked whether they would use a specific AI system. They are asked whether AI is good or bad for humanity. That wording invites a broad moral and political judgment. It makes the article less about adoption intention in the narrow sense and more about how AI is placed within respondents' wider views of progress, governance, and social order.

The concept of *governed modernity* summarizes this shift, but it is used as a sensitizing concept rather than a directly measured latent construct. "Governed" concerns whether states, employers, experts, norms, and public institutions appear capable of directing technological change legitimately and tolerably. "Modernity" concerns AI's symbolic association with scientific development, economic transformation, social reorganization, and collective futures. The construct therefore draws attention to whether respondents expect AI-led change to remain institutionally controllable, distributively acceptable, and compatible with valued forms of human agency.

This framing is related to the literature on sociotechnical imaginaries. Jasanoff and Kim (2015) describe collectively held visions of desirable futures animated by science and technology and connected to public purpose and social order. Sartori and Bocca (2023) apply a sociotechnical perspective to public understandings of AI, emphasizing that awareness, knowledge, narratives, institutions, and emotional responses jointly shape how AI is imagined. A single broad AI evaluation cannot recover a complete sociotechnical imaginary. The relevance of this literature is instead that the wide range of predictors associated with AI evaluation indicates that respondents locate AI inside broader expectations about authority, progress, fairness, and human control. The article's contribution is thus not the general claim that AI attitudes extend beyond technology acceptance, which is already established, but the identification of a comparative value architecture through which societal evaluation is organized.

## 2.2 Ambivalence, algorithmic authority, and weak attitude sorting

Research on general attitudes toward AI increasingly shows that such attitudes are not one-dimensional. Schepman and Rodway (2020) introduced the General Attitudes toward Artificial Intelligence Scale (GAAIS), demonstrating that positive and negative attitudes toward AI can coexist rather than simply mirror one another. Their later confirmatory work links positive AI attitudes to general trust and negative attitudes to corporate distrust (Schepman and Rodway 2023). Stein et al. (2024) similarly argue that broad AI attitudes are measurable as a meaningful general disposition and show that these attitudes connect to demographic and personality differences. Shum and Lau's (2024) latent profile analysis further underscores that ambivalence is common: many respondents are neither clear enthusiasts nor outright skeptics.

A related literature on algorithmic judgment helps explain why public evaluations may be conflicted. Dietvorst, Simmons, and Massey (2015) show that people can become averse to algorithms after seeing them err, even when algorithms outperform humans on average. Yet other work documents "algorithm appreciation" in some settings, especially when algorithms are seen as more capable or objective than human decision-makers (Logg, Minson, and Moore 2019). Castelo, Bos, and Lehmann (2019) show that aversion is task-dependent and stronger when tasks are felt to require intuition or human sensitivity. Longoni, Bonezzi, and Morewedge (2019) make a related argument in healthcare, where resistance to medical AI is shaped by the perception that AI neglects patient uniqueness.

These studies are important because they suggest that public reactions to AI are not only about performance. They are also about the legitimacy of delegated authority and the boundaries of what should remain recognizably human. The present article extends that insight from domain-specific experiments to a broad comparative values survey. It asks whether general attitudes toward AI are anchored in wider beliefs about agency, voice, fairness, moral order, and institutional legitimacy.

Person-centered methods provide a complementary way to study this heterogeneity. Variable-centered analysis asks which variables are associated across a population, whereas person-centered analysis identifies configurations that recur among people (Bergman and Trost 2006; Woo et al. 2024). Existing person-centered AI research usually constructs profiles partly from AI-specific evaluations. Shum and Lau's (2024) enthusiast, skeptic, and indecisive profiles are useful descriptions of ambivalence, but separation on AI attitudes is expected because AI items define the profiles. The stronger question for the present article is whether profiles formed from broader values, without AI attitudes, naturally reproduce those camps.

The structure of mass belief systems provides a basis for expecting only partial sorting. Converse (1964) argued that mass attitudes are often less ideologically constrained than elite belief systems. Recent large-scale evidence similarly finds that citizens' issue positions are fragmented and multidimensional rather than consistently organized along one ideological axis (Mongrain and Walgrave 2026). Coherence becomes more likely when issues are linked to salient groups, party competition, or sustained elite cues (Elder and O'Brian 2022). AI is consequential across science, work, authority, surveillance, national development, and social protection, but those implications need not yet be consolidated into one ideological package across 24 national settings.

This literature motivates the article's description of generalized AI opinion as **value-embedded and contextually differentiated**. Broad values organize AI evaluation through recurring social configurations, while each configuration contains both favorable and unfavorable evaluators. The distinction is between *defining values*, which make a respondent profile different from other profiles, and *discriminating values*, which distinguish favorable from unfavorable evaluators within that profile. A profile may be strongly secular without secularism explaining its internal AI divide, while less visually distinctive differences in legitimacy or socioeconomic and educational resources may be highly discriminating. The profiles therefore locate the social contexts in which particular values become consequential.

## 2.3 From generalized trust to a source-specific trust architecture

Trust has become one of the central concepts in scholarship on AI governance. Chen and Wen (2021) show that public trust in AI is tied to attitudes toward governments and corporations. Choung, David, and Ross (2023) argue that trust in AI is inseparable from ethical concerns about fairness and institutional responsibility. David, Choung, and Seberger (2024) similarly show that views about who should govern AI are structured by trust and ethics, while Lahusen, Maggetti, and Slavkovik (2024) emphasize that trust in AI governance is best understood as a socio-technical problem involving institutions, regulatory systems, and machine systems at once. At the same time, publics often support stronger oversight even when they recognize AI's potential benefits (Nussberger et al. 2022; Vesely and Kim 2024).

This literature often implies a trust-transfer model: publics that trust institutions are more likely to trust AI. That expectation is plausible and often correct. But it is not the only possible pathway. Molina and Sundar (2024) show that distrust in humans can, under some conditions, increase trust in AI. The possibility matters greatly for comparative public opinion. In some settings, AI may be welcomed because institutions are trusted to steer it responsibly. In others, AI may be welcomed because existing institutions are viewed as corrupt, biased, or ineffective and machine systems are imagined as a cleaner alternative.

Recent governance work also underscores that these questions operate across levels of authority rather than only at the level of a single institution. Choung, David, and Seberger (2024) argue for a multilevel framework for AI governance that links public trust, institutional responsibility, and governance architecture. That perspective is especially useful here because many WVS predictors speak not only to whether institutions are trusted, but also to whether respondents regard delegated authority as legitimate in the first place.

The category of trust should consequently be disaggregated by source and object. Trust in an AI system's competence, technology companies, government regulators, scientific experts, news organizations, and social-media platforms are not interchangeable. Busuioc and Maggetti (2026) similarly argue against treating trust maximization as an unqualified governance objective and instead emphasize calibrated trust grounded in vulnerability, accountability, and demonstrated trustworthiness. This distinction helps explain why institutional legitimacy can recur in selected contexts while trust in information intermediaries is still more contingent and may attenuate or reverse after adjacent constructs enter jointly.

Two candidate pathways follow. A *stewardship pathway* links favorable evaluation to confidence that legitimate institutions can govern AI responsibly. A *substitution pathway* makes AI attractive relative to distrusted human intermediaries; Molina and Sundar's (2024) content-moderation evidence directly demonstrates this possibility in one domain. The present WVS analysis cannot identify either mechanism because it does not ask whether respondents want AI to replace a particular institution. It can, however, test whether legitimacy and intermediary confidence occupy distinct positions in the conditional architecture. The resulting theoretical claim concerns a source-specific trust architecture, not trust in the abstract.

## 2.4 Multiple routes through modernity

A smaller but especially relevant body of work treats attitudes toward AI as embedded in broader value systems. O'Shaughnessy et al. (2023) show that individualism, egalitarianism, risk aversion, and techno-skepticism shape attitudes toward AI adoption and governance. Wang's (2025) cross-national analysis of twenty countries likewise demonstrates that both individual-level and country-level factors matter for AI perceptions. These findings align with broader theories of modernization and value change. Inglehart (1997) and Inglehart and Welzel (2005) argue that values related to secularization, self-expression, autonomy, and authority vary systematically with social and economic development. If AI is interpreted as part of a broader project of modernity, it should be entangled with these wider values rather than treated as a stand-alone technological preference.

But modernization theory alone is not enough. The same values may push attitudes toward AI in different directions depending on how AI is interpreted. Emancipative and secular values may support AI when it is read as openness, progress, and rational problem-solving. The same values may produce skepticism when AI is interpreted as opaque delegation, surveillance, or centralization of power. Traditionalism may produce skepticism when AI is seen as a threat to moral order or labor dignity. Yet traditionalism can also coexist with AI optimism when AI is placed within narratives of national strength, social discipline, or competent hierarchy. The theoretical task, then, is not simply to place AI attitudes on a single modern-traditional axis, but to recover the combinations through which values become meaningful.

Work on worldview predictors of scientific innovation pushes in a similar direction. Veckalov et al. (2023) show that attitudes toward AI, nanotechnology, and human gene editing are structured by broader worldview differences, including value orientations associated with scientific skepticism and cultural outlook. Their findings are useful for the present study because they suggest that AI attitudes should be read as part of a wider syndrome of reactions to scientific innovation rather than as an isolated technology-specific preference.

Recent meta-analytic evidence similarly finds that national cultural dimensions moderate several antecedents of AI-adoption attitudes and intentions (Wang et al. 2026). That result establishes the importance of cultural contingency but leaves two questions open for broader societal evaluation: whether national averages conceal recurring value configurations within countries, and whether the same value has a comparable association across those configurations. The present analysis addresses these questions without treating a country as culturally homogeneous or equating a judgment about AI's consequences for humanity with intention to use a specific system.

The theory of multiple modernities provides a more suitable comparative language than a single modern-traditional continuum. Eisenstadt (2000) argues that modernity is reconstructed through different cultural and institutional programs rather than diffusing as one uniform Western model. Scientific development, national strength, bureaucratic competence, religious authority, capitalism, social protection, and individual autonomy can therefore be recombined. AI is especially open to such recombination because it can symbolize emancipation, expert capacity, economic catch-up, sovereign power, orderly administration, or technocratic domination.

The relevant dimensions should thus be analytically decomposed. At least five routes may connect broader modernity to AI evaluation: confidence in science and technical progress; confidence in institutional capacity; education and material resources for adaptation; openness and information access; and democratic or humanistic concern about opaque delegation, surveillance, concentrated power, and loss of human control. Comparative evidence that trust and acceptance can be higher in many emerging economies than in advanced economies further cautions against treating Western secular-liberal publics as the universal model of AI optimism (Gillespie et al. 2025). The article uses *multiple routes through governed modernity* to describe these combinations without claiming that any one route causes favorable evaluation.

### 2.5 Common anchors across heterogeneous contexts

Existing scholarship gives us important pieces of the puzzle. General attitudes toward AI are measurable and often ambivalent; trust, perceived risk, education, age, and cultural values matter; and national context shapes how technological change is perceived. What is less developed is a way to distinguish three propositions that are often treated as interchangeable. A feature may be globally important in a predictive model. It may be associated with different average attitudes across social profiles. Or its association with AI may change depending on the surrounding value context. These are not the same empirical claim.

Recent comparative evidence identifies both recurring correlates and meaningful cross-economy differentiation. Rice and Wu (2025), analyzing 32,330 respondents across twenty economies, identify awareness, education, religiosity, support for science, science-news use, institutional trust, and economy-level differences, while finding mostly small individual effects and meaningful cross-economy variation. Such findings motivate a more precise vocabulary. **Directional portability** asks whether a predictor points the same way across settings. **Salience portability** asks whether its rank or importance is similar. **Conditional portability** asks whether it retains its association once adjacent values enter jointly. **Interpretive portability** asks whether the same mechanism is plausible in each context. A predictor can travel in one sense and not another.

The distinction matters because a globally recurrent correlate need not represent one portable mechanism. Scientific optimism may travel widely because AI is commonly understood as a technological project. Other values are more relational. Human agency can support AI when it reflects distance from inherited authority, but it can support skepticism when AI is read as opaque delegation. Egalitarianism can align with optimism when AI is imagined as expanding access, or with opposition when it is associated with market concentration and displacement. Secular reasoning may differentiate technological openness inside a religious profile while having little effect inside an already secular profile. The surrounding value configuration can therefore alter the salience and even the direction of an association.

This creates a methodological risk for person-centered analysis. If AI attitudes are included in clustering, upweighted, or used to select and label clusters, then outcome-favorable groups are partly guaranteed by construction. Such clusters may still be descriptively useful, but they cannot independently establish that broader values naturally organize themselves around AI evaluation. A stronger test forms profiles without the outcome, freezes those assignments, and then asks whether average attitudes and within-profile predictors differ afterward. This design shifts attention from outcome purity to conditional heterogeneity.

The comparative scope remains important. Research on AI attitudes is concentrated in a limited set of countries, even though labor-market exposure, institutional context, and national discourse all shape technological evaluation. Wang (2025) provides broad cross-national evidence on AI perceptions, while OECD work by Lane and Saint-Martin (2021), Georgieff and Hyee (2021), and Georgieff (2024) emphasizes employment restructuring and wage inequality. The present study adds a complementary question: which correlates appear portable across value profiles, and which acquire different meanings across those profiles even within a common pooled survey environment?

Comparative-methods research provides a general rationale for treating slope similarity as an empirical question. Heisig, Schaeffer, and Giesecke (2017) show that assuming invariant lower-level relationships across clusters can distort contextual inference. The present analysis is descriptive rather than causal, but the same principle applies: national and respondent contexts may differ in slopes as well as intercepts. The expected pattern is neither complete universality nor wholly separate national regimes, but **common anchors in plural configurations**. Science optimism is the clearest recurring anchor; legitimacy, socioeconomic and educational resources, and other values vary more in rank, conditional role, and surrounding social meaning.

## 2.6 Explicit non-awareness and evaluability

Exposure to AI discourse is theoretically prior to a substantive evaluation. Diffusion research and comparative AI surveys therefore distinguish familiarity or awareness from the direction of opinion (Rice and Wu 2025; Sartori and Bocca 2023). Yet the WVS item supports only a narrow version of that distinction. The AI attitude question includes an explicit "not heard of AI" option, but it also contains "don't know," nonresponse, and availability codes. A respondent who answers "don't know" may have heard of AI, while the absence of a "not heard" response in a country may reflect instrument or translation differences rather than universal familiarity.

The primary outcome in this article is consequently favorable versus unfavorable evaluation among respondents selecting those two substantive responses. A secondary **evaluability** contrast compares substantive evaluators with respondents explicitly coded "not heard" among the 0/1/2 response set. It is not labeled a general awareness measure, and alternative treatments of "don't know" are reported as comparability diagnostics rather than interchangeable outcomes. This narrower language preserves the useful distinction between entry into an issue public and evaluative positioning without claiming more than the question measures.

## 2.7 Outcome separation and layered inference

The design follows the distinction between variable-centered and person-centered questions (Bergman and Trost 2006; Woo et al. 2024). Predictive modeling identifies a broad architecture of correlates; clustering identifies recurring configurations of respondents or countries; conditional models then ask which thematic dimensions distinguish evaluations inside fixed contexts. None of these operations substitutes for the others.

Excluding AI attitudes from profile formation follows the same design principle that motivates separating class indicators from distal outcomes in latent class and profile analysis (Nylund-Gibson, Grimm, and Masyn 2019; Bakk and Kuha 2021). The present procedure is not a latent-class model and does not correct for classification error. Its contribution is to form contexts without the outcome, freeze them, and then examine outcome distributions and conditional associations. This design reveals whether broad value configurations organize mean evaluations, within-profile differentiation, or both.

# 3. Data and Analytic Strategy

## 3.1 Data, outcome, and estimands

The analysis uses provisional World Values Survey (WVS) Wave 8 data available as of April 2026 (World Values Survey Association 2026).[1] The 40,960 respondents come from 24 country samples: Andorra, Bangladesh, Bolivia, Canada, Cote d'Ivoire, Colombia, Egypt, Greece, Guatemala, Indonesia, Japan, Kenya, Morocco, Malaysia, Nigeria, Pakistan, Peru, Singapore, Slovakia, Sweden, Taiwan, Thailand, the United States, and South Africa. These samples are not treated as a probability sample of the world.

Q121 asks whether artificial intelligence is good or bad for humanity. The source states are 0 = not heard of AI, 1 = unfavorable, 2 = favorable, -1 = don't know, -2 = no answer, and -4/-5 = not asked or unavailable.[2] The primary binary outcome compares the 19,389 favorable and 14,414 unfavorable responses (33,803 evaluators). The full sample also contains 4,730 explicit not-heard responses, 1,963 don't-know responses, 392 no answers, and 72 not-asked or unavailable records. "Support" and "opposition" are avoided as labels because Q121 asks for a societal evaluation, not behavioral or policy support.

Descriptive estimates use the WVS respondent weight. Equal-country estimates rescale those weights so each of the 24 samples contributes the same total; their target is the average of these observed country samples, not a global population parameter.[3] The prediction and joint association models use equal-country WVS weights. The source includes a primary-sampling-unit identifier in all 24 samples but no harmonized stratum field. Country-delete jackknife inference addresses dependence at the country-sample level; it is not a full reconstruction of each national survey design.

The source-labeled sex variable (Q228) records responses to a gender-description question, not a harmonized measure of sex assigned at birth. The provisional source lists woman, man, transgender, non-binary, other, and prefer not to say; some national samples use an older scale. Prediction models encode the first five responses as unordered categories, treat prefer not to say as nonresponse, and omit a country-relative copy of this field. Appendix A3 reports the coding sensitivity.

## 3.2 Predictor taxonomy and thematic dimensions

WVS items relevant to AI are dispersed across science, institutions, resources, authority, distribution, information, and social boundaries. The analysis uses two complementary maps. A ten-class taxonomy organizes item-level predictors from the pooled model (Table 1). Eleven thematic measures provide common variables for profile description and joint within-context models (Table 2). The taxonomy preserves question specificity; the thematic measures summarize selected, directionally aligned content. Neither is presented as a validated reflective measurement model.

---

[1] The exact provisional source refresh is `WVS8_v1_Spss_eng_v20260421`.

[2] The source-native April 2026 coding is used throughout. Records outside 0/1/2 are retained in the response-state audit but excluded from favorable-versus-unfavorable models.

[3] Equal-country weighting prevents larger national samples from dominating. It does not make the 24-country set globally representative.

*Table 1. Literature-grounded taxonomy of item-level predictors.*

| Predictor taxonomy | Relevance to generalized AI evaluation |
|---|---|
| Scientific and technological progress | AI as beneficial progress or suspect innovation |
| Social location, life course, and human capital | Exposure, education, resources, and labor-market position |
| Institutional legitimacy and governance trust | Confidence that institutions can govern AI |
| Democratic voice, autonomy, and accountability | Human control, political voice, and legitimate delegation |
| Economic fairness and distributive order | Distribution of technological gains and losses |
| Emancipative and social-liberal values | Choice, equality, openness, and principled skepticism |
| Traditionalism, religiosity, and moral authority | Inherited authority, religion, family, and moral order |
| Security, order, and punitive authority | AI as capacity, control, risk, or disruption |
| Media and information intermediaries | Exposure to and confidence in mediated knowledge |
| Identity, migration, and territorial embeddedness | Openness, belonging, and boundary making |

The many-to-many mapping from taxonomy to measures was developed from the literature and then audited against item wording, response direction, all-country availability, and empirical coherence. For example, emancipative and traditional-authority content separates into personal choice, human agency, secular reasoning, and coercive order; media content separates into information connectedness and intermediary trust. Distinctive predictors such as preference for expert decision making remain item-level findings rather than being forced into a measure.

*Table 2. Thematic measures used for profile description and joint models.*

| Thematic measure | Items | Higher-score meaning |
|---|---|---|
| Science-and-technology optimism | 1 | Science and technology make the world better rather than worse (Q118) |
| Institutional legitimacy | 43 | Confidence in institutional competence and integrity; lower perceived corruption and electoral unfairness |
| Structural advantage | 12 | Greater educational, economic, professional, and locational resources; it is not a direct measure of digital skill or AI readiness |
| Information connectedness | 8 | More regular use of news, internet, email, mobile, and interpersonal information channels |
| Personal-choice liberalism | 27 | Greater choice and equality in reproductive, sexual, marital, family, and gender domains |
| Human agency and participation | 41 | Autonomy, efficacy, democratic voice, civic membership, and political participation |
| Coercive order | 20 | Greater priority for authority, obedience, security, punitive enforcement, and force |
| Trust in information intermediaries | 6 | Greater confidence in press, broadcast media, experts, and related intermediaries |
| Redistributive protection | 4 | Stronger support for equality, taxation/subsidy, unemployment protection, and state responsibility |
| Boundary openness | 8 | Greater acceptance across nationality, religion, race, and language and more favorable immigration views |

| Thematic measure | Items | Higher-score meaning |
|---|---|---|
| Secular-individual reasoning | 22 | Greater distance from religious authority and inherited family or moral obligations |

The 192 retained items form formative summaries: the indicators collectively define a domain and are not assumed to be interchangeable effects of one latent trait. Q118 remains a single item because the discarded alternative, whether policy should follow scientific evidence, is a distinct judgment. Multi-item alpha values range from .503 for redistributive protection to .899 for institutional legitimacy; these are diagnostics rather than validity tests. The appendix describes the construction and audit; the research-materials supplement supplies the complete machine-readable item map.

For regression, each measure is constructed once from absolute item responses, centered on its survey-weighted country mean, and divided by one equal-country pooled within-country standard deviation (SD; Enders and Tofighi 2007; Bell and Jones 2015). A coefficient therefore compares respondents separated by the same common within-country distance in every profile. Absolute and country-relative copies of the same measure are not entered as separate causes.

### 3.3 Variable-centered prediction

Extreme gradient boosting (XGBoost) models favorable versus unfavorable evaluation among the 33,803 evaluators (Chen and Guestrin 2016). The clean pooled builder retains 287 source fields and excludes country and region identifiers, subnational and birthplace fields, language, ethnicity, country-specific education fields, party codes, all Q79 regional-organization items, and high-cardinality administrative text. Preprocessing is learned inside each training fold: numeric fields receive training-fold median imputation and nominal fields are one-hot encoded from training categories. Eligible ordered fields also receive within-country standardized copies calculated from training-fold country moments. No country indicator is included.

The headline model uses pooled respondent splits stratified by country and outcome. Every country appears in training, validation, and test data; the design therefore measures interpolation to new respondents in observed national settings, not transport to a new country. Three 64/16/20 train/validation/test repetitions are used for performance and ablations. TreeSHAP, the tree-model implementation of Shapley additive explanations (SHAP), describes fitted predictive contributions on held-out respondents (Lundberg and Lee 2017). The area under the receiver operating characteristic curve (AUC) comparison between the full model and an absolute-only model assesses incremental discrimination from country-relative copies. The proportion of SHAP assigned to either block is treated as representation-dependent, not as an identified within/between decomposition.

A duplicate-predictor sensitivity removes 35 derived indices and recodes whose component items remain in the model. Separately fitted national models use absolute fields and ten stratified respondent splits per country. Their main summary is how often a source enters the top ten and with what direction; a one-shot rank heatmap is supplementary. Finally, leave-one-country-out (LOCO) models train on 23 countries and predict the omitted sample using absolute fields. LOCO tests portability, not the omitted country’s own predictive structure.

### 3.4 Outcome-independent respondent profiles

The profiles locate recurring configurations of values and social position within which AI-evaluation associations can be compared. They are not optimized to separate favorable from unfavorable respondents. Q121 is absent from representation fitting, distances, the cluster-count scan, and profile labels.

The corrected representation retains 243 harmonized non-AI sources. It excludes all country, origin, region, language, ethnicity, Q79 regional-organization, and country-specific education fields. A source must also have at least 20 substantive observations in every included country, preventing questionnaire availability from serving as a country fingerprint. Missing responses are placed at the standardized mean for ordered variables or the observed category-proportion centroid for nominal variables; missingness indicators are not included. Country-relative copies are added for 38 eligible sources, and absolute and relative blocks receive equal aggregate scale. Expansion produces 1,280 engineered features.

Pairwise controlled manifold approximation projection (PaCMAP) reduces the representation to ten dimensions using 50 embedding neighbors (Wang et al. 2021). A symmetrized 25-nearest-neighbor graph then constrains Ward agglomeration to local merges (Ward 1963). Candidate cuts from 10 through 15 profiles are compared without Q121 using Ward merge discontinuity, silhouette, Davies-Bouldin fit, stability across 15-, 25-, and 50-neighbor connectivity graphs, and size balance. K=11 has the strongest median Ward jump among the leading cuts (1.122), the best median silhouette (.232), a Davies-Bouldin value of 1.280, and mean graph-stability adjusted Rand index (ARI) of .659. Its profiles range from 4.9% to 18.4% of the sample; all 40,960 respondents are assigned and none falls below the prespecified 2% reporting floor.

Three corrected alternative K=11 partitions vary the PaCMAP seed or graph neighborhood. Their ARI with the primary solution ranges from .430 to .678. The legacy pre-audit partition has ARI .244 with the corrected solution. Profile-specific inference is therefore conditional on a data-adaptive representation. The article reports whether conclusions recur across corrected partitions rather than treating the selected labels as fixed population kinds.

Country composition is tested separately by sampling 1,000 respondents per country before preprocessing and PaCMAP fitting. Two balanced reconstructions have ARI .425-.431 with the primary labels on the same respondents. Their cluster shares remain 3.3%-21.9%, so the sensitivity does not recreate tiny residuals, but it confirms that exact membership depends partly on national sample composition.

### 3.5 Within-profile associations

After assignments are frozen, Q121 is attached for descriptive distributions. A weighted variance decomposition measures the share of favorable/unfavorable variation between profile means. Conditional associations are estimated in one logistic model containing country fixed intercepts, profile indicators, all eleven thematic measures jointly, and every profile-by-measure interaction. These interactions yield one set of slopes for each profile, pooling its respondents across the national samples. All 11 profiles enter the same model, yielding 33,045 evaluators with complete constructed scores.

Country-delete jackknife refits omit each of the 24 samples in turn (MacKinnon, Nielsen, and Webb 2025). Profile-specific intervals use a t reference with 23 degrees of freedom. Benjamini-Hochberg (BH) correction (Benjamini and Hochberg 1995) is applied to the 11 omnibus profile-heterogeneity tests and separately to the 121 profile-measure coefficients. Average marginal effects and adjusted 10th-to-90th-percentile (P10-P90) probability contrasts translate log-odds estimates to the probability scale (Mize 2019). Diagnostics flag limited within-profile range, sparse tails, insufficient component variation, high variance inflation factors (VIF), and poor condition number; no primary cell is flagged.

These p values are conditional on the selected profile construction. They are not confirmatory post-selection inference. The primary safeguard is substantive recurrence across the three corrected partitions, each refitted with a full 24-country delete jackknife. A regularized model allowing both country- and profile-varying slopes is an additional point-estimate sensitivity. Cluster descriptions and item-level screens are generated deterministically from the survey and model tables. Numerical and inferential claims are based on these retained results; AI assistance with method development and interpretation is described in Section 3.8.

### 3.6 Outcome-independent country contexts

Countries are represented by their weighted shares across C0-C10 and their weighted means on the eleven absolute thematic measures. An equal-weight distance combines Hellinger distance over profile shares with Euclidean distance over standardized axis means. Average-linkage cuts from K=3 through K=6 are assessed using silhouette, merge interval, minimum group size, respondent-bootstrap stability, partitioning around medoids (PAM), alternative axis/profile weights, and profile-only and axis-only solutions. The primary K=3 solution has no singleton and contains 4, 10, and 10 countries. The three contexts are descriptive comparative strata, not national latent classes or development stages.

The same joint, country-centered thematic model is then estimated separately for (a) favorable versus unfavorable evaluation and (b) the narrow evaluability contrast among Q121 responses 0, 1, and 2. All 24 countries enter the country-delete inference. Because the latter excludes don't-know and nonresponse states and two countries record no explicit not-heard responses, it is interpreted only as a secondary instrument-dependent diagnostic.

### 3.7 Comparability, multiplicity, and scope

The response-state audit reports every Q121 code by country, fieldwork period, collection mode, effective sample, and first-stage coding sensitivity. Seven countries have at least 10% indeterminate or nonresponse states, Morocco and Singapore record no explicit not-heard responses, and treating don't know as nonevaluation changes the first-stage estimate by as much as 32.4 percentage points. The WVS source provides one shared variable label and codebook but cannot establish equivalence of national translations or routing.

For thematic measures, 170 of 192 items have at least 20 valid observations in every country; country-axis score coverage ranges from 82.1% to 100% under the component threshold. Fieldwork began in 2024 for 16 samples and 2025 for eight; the country-level correlation between timing and favorable evaluation is -.082 and cannot be separated from country in this cross-section.

Two comparability sensitivities refit the complete joint models while holding profile and context assignments fixed. The first retains only the 170 components observed adequately in all 24 countries. The second retains the 16 samples whose fieldwork began in 2024. These checks assess coefficient and decision stability; they do not establish translation invariance or identify a time effect.

The analysis was developed iteratively and is best described as exploratory and abductive. Multiplicity correction limits the most direct families of tests, but it does not erase researcher degrees of freedom in representation, profile construction, or interpretation. Accordingly, effect recurrence, sensitivity, observed covariate range, and exact estimands receive more weight than isolated significance.

### 3.8 AI-assisted research workflow

The project used substantial agentic AI assistance in research development, not only language editing. OpenAI Codex and ChatGPT assisted with methodological ideation, writing and debugging preprocessing and analysis code, producing visualizations, and drafting and revising the manuscript. Earlier exploratory stages used GLM-5.2 (`z-ai/glm-5.2`) through API calls to summarize clusters, predictors, and thematic associations. These suggestions informed method development and interpretation. Reported quantities are computed from the WVS data by retained scripts, rather than supplied as numerical estimates by a language model. Research decisions and responsibility remained with the authors, who reviewed and revised AI-assisted material. The appendix distinguishes exploratory AI interpretations from the publication analysis.

## 4. Cross-Country Predictive Regularities

### 4.1 Pooled prediction

The pooled respondent-split model reaches mean equal-country AUC .769 (SD .0015). The absolute-only model reaches .760, so country-relative copies add about .009 AUC. Removing Q118 lowers AUC to .747; removing the broader science/technology family produces .746. A model without 35 duplicate indices and recodes reaches .768, only .0014 below the primary model, and produces similar ablation losses. The predictive architecture is therefore not an artifact of duplicated derived fields.

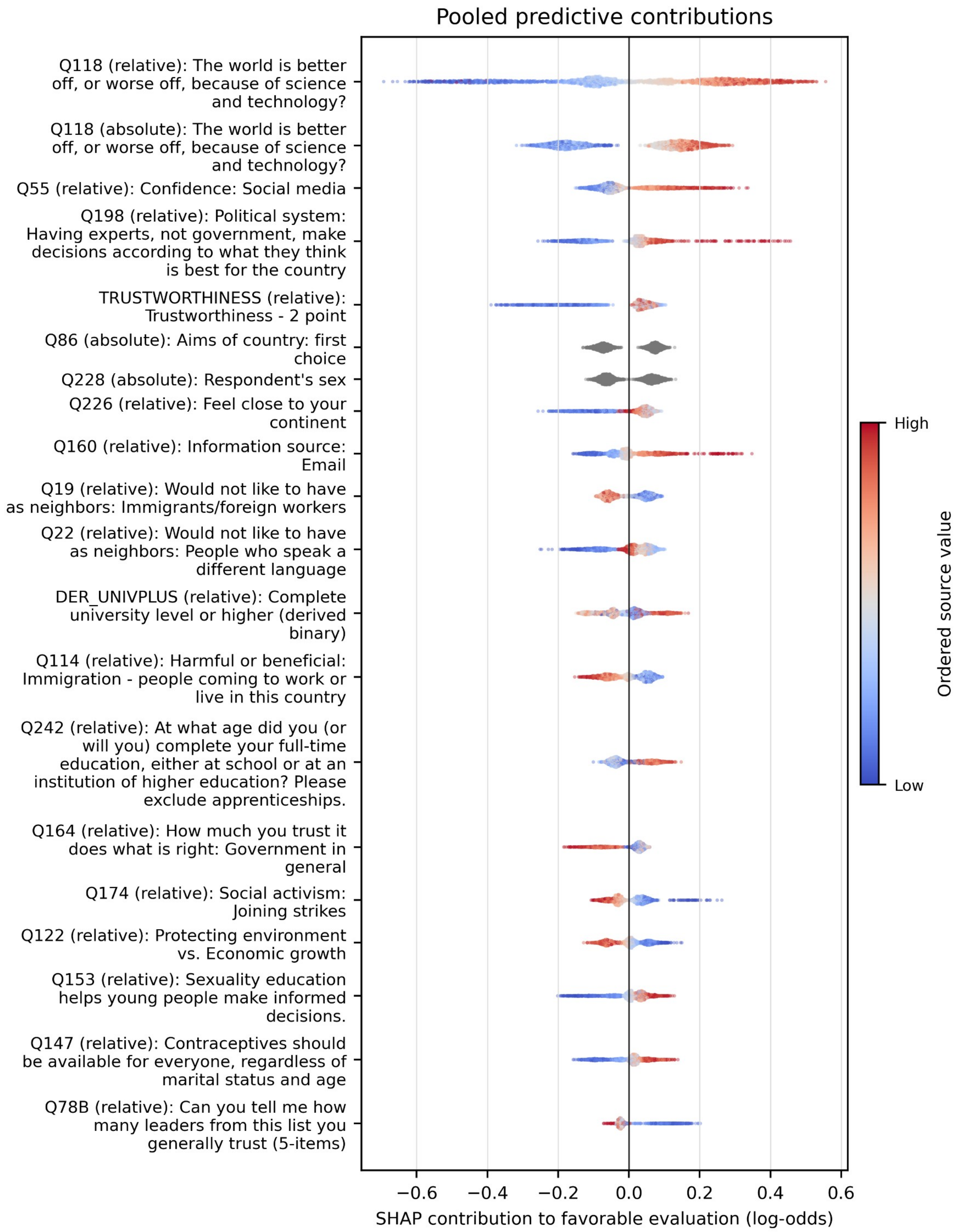

*Figure 1. Signed source-level TreeSHAP distributions from the seed-42 pooled respondent-split test set. Positive values increase predicted favorable evaluation. Color orders only sources with a defensible low-to-high coding; nominal sources are gray. Absolute and country-relative variants are shown separately. Plotting code was developed with AI assistance.*

Q118 has by far the widest signed contribution. Respondents who see science and technology as improving the world receive more favorable AI predictions, both absolutely and relative to compatriots. The relationship is theoretically meaningful rather than tautological: Q118 is a broad judgment about scientific and technological progress, while Q121 evaluates AI specifically.

Secondary contributions supply more concrete hypotheses. Favorable predictions are associated with receptivity to expert decision making, university education or later completion of education, email use, immigration openness, continental attachment, universal contraceptive access, support for sexuality education, interpersonal trustworthiness, and confidence in social media. Some institutional items point the other way after conditioning on the rest of the model: relative confidence in government and confidence in more political leaders can contribute toward unfavorable predictions. These are predictive contributions among correlated fields, not independent causal effects. The differentiated signs also show why neither "trust" nor "liberalism" is an adequate one-variable account.

The model allocates much of its SHAP magnitude to country-relative variants, but that allocation is not interpreted as a causal or contextual effect. Absolute and relative copies encode overlapping information, and correlated features can exchange attribution. The defensible finding is the modest out-of-sample gain from adding relative position, not a numerical partition of social causation.

## 4.2 National recurrence and portability

Separately fitted national models average AUC .731 (median .731; country range .535-.836). Their median pairwise top-10 Jaccard similarity across ten splits is only .271. Most national rank differences are therefore too unstable to support a detailed country-by-country hierarchy.

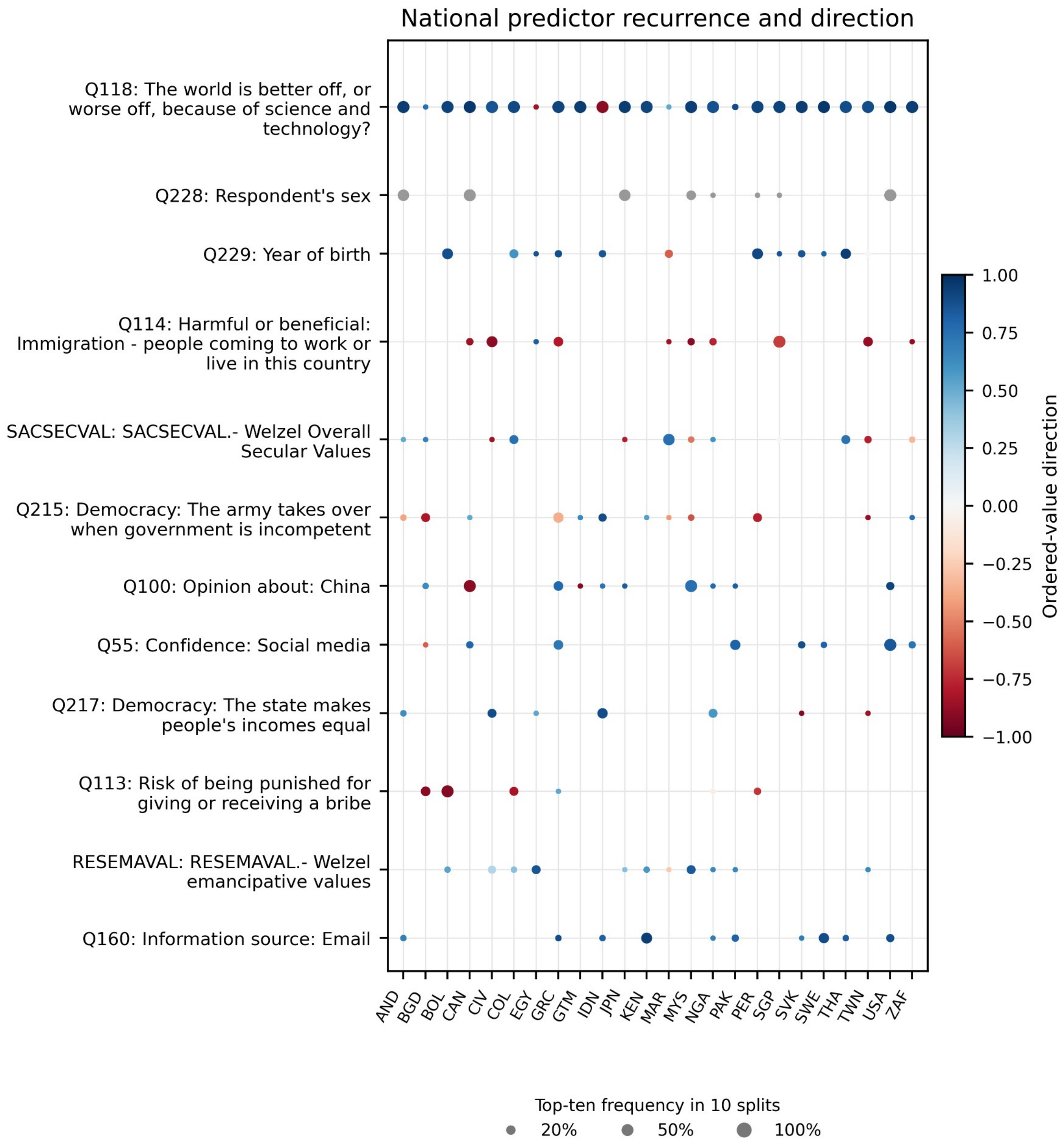

*Figure 2. Recurrence and direction of leading sources in ten separately fitted models per country. Marker size records the frequency with which a source enters that country's top ten; color summarizes fitted direction where an ordering is defensible. Gray markers indicate unordered categories. Empty cells indicate no top-ten appearance. Plotting code was developed with AI assistance.*

One regularity is unusually strong. Q118 appears in at least one split's top ten in all 24 countries and in at least half the splits in 20. The next most recurrent source, nominal gender-description categories (Q228), meets the half-of-splits threshold in five countries and appears at least once in eight; its unordered contribution is shown without a directional color. Birth year and immigration evaluation each meet the threshold in four countries. Secular values, social-media confidence, military rule, income equality, email use, views of China, and perceived bribery risk recur in smaller subsets. Figure 2 therefore supports a common science-optimism anchor surrounded by much less stable national feature combinations.

LOCO asks a different question: whether a model trained outside a country transports to it. Macro country AUC is .654 (median .673; range .482-.747), with Bangladesh, Egypt, Indonesia, and Morocco below .55. These results do not describe what determines evaluations within those countries. They show that relationships learned across the other 23 samples transfer unevenly. Detailed LOCO metrics and the transported-model SHAP display are reported in the appendix.

# 5. Outcome-Independent Respondent Profiles

## 5.1 Profile composition and AI evaluation

Figure 3 locates the profiles on the same 11 thematic measures later used in the joint outcome model. Table 3 summarizes their composition and favorable evaluation. These descriptive positions define the profiles; they should not be confused with within-profile associations with AI.

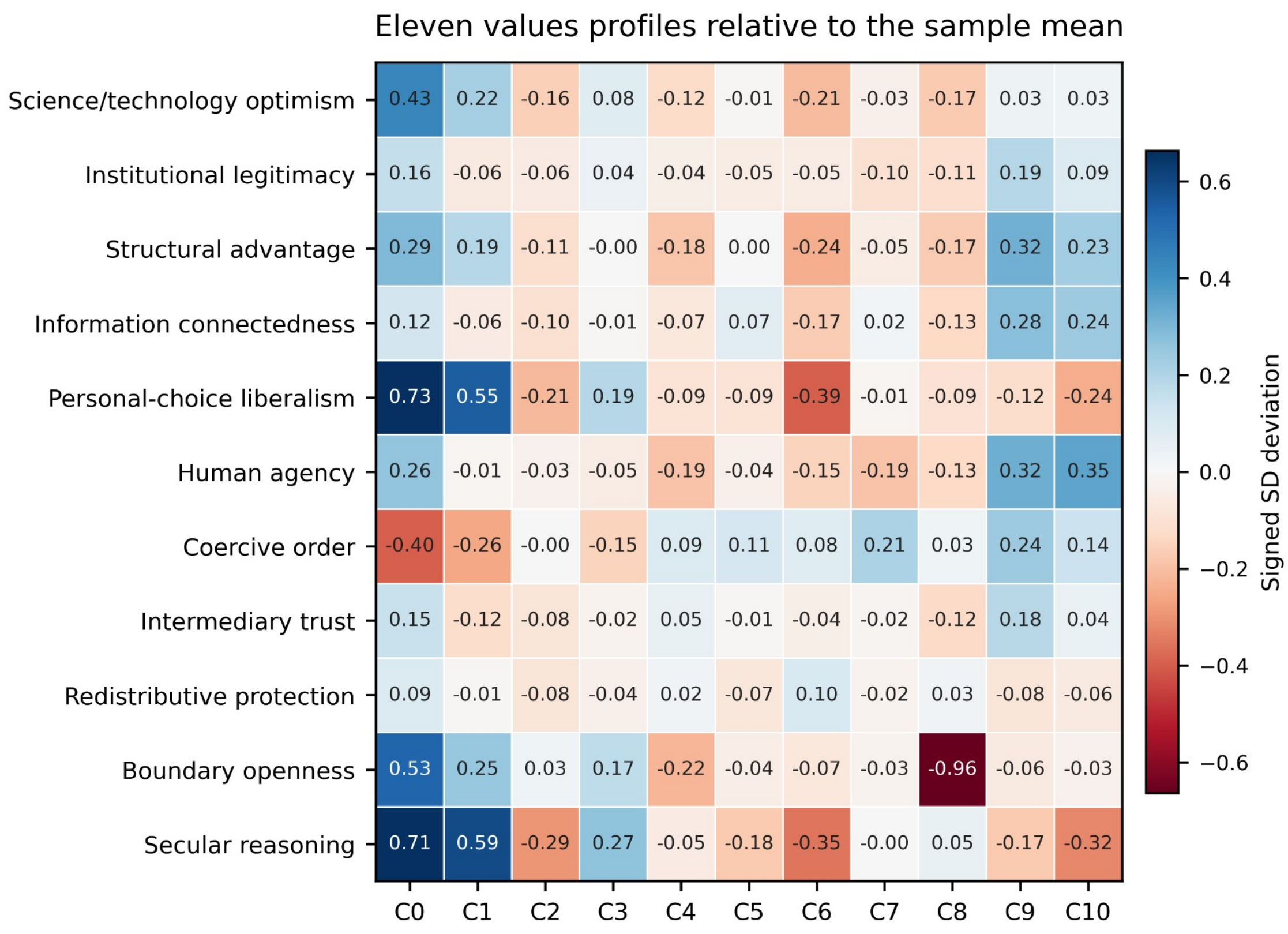

*Figure 3. Profile positions on the audited thematic measures relative to the full 24-sample mean. Q121 was not used in profile formation. Plotting code was developed with AI assistance.*

*Table 3. Outcome-independent respondent profiles and favorable evaluation.*

| Profile | Share | Favorable among evaluators | Brief characterization |
|---|---|---|---|
| C0 Secular cosmopolitan choice liberals | 9.9% | 54.1% | Highly secular, choice-liberal, and boundary-open; concentrated partly in Canada and Sweden |
| C1 Secular private-choice individualists | 9.3% | 62.2% | Secular and liberal in intimate choice, with weaker institutional and intermediary confidence |

| Profile | Share | Favorable among evaluators | Brief characterization |
|---|---|---|---|
| C2 Religious social traditionalists | 8.8% | 49.1% | Religious and restrictive in family morality, but not strongly coercive or boundary-closed |
| C3 Secular moderate pluralists | 8.8% | 55.2% | Moderately secular, choice-liberal, and open without C0/C1's extremes |
| C4 Disadvantaged low-participation order seekers | 8.3% | 53.1% | Lower structural advantage, participation, and openness with a modest order orientation |
| C5 Faith-centered institutional mainstream | 12.7% | 60.6% | Religious mainstream combining modest order, information connection, and institutional incorporation |
| C6 Religious restrictive material solidarists | 18.4% | 60.3% | Restrictive personal morality and lower structural resources alongside redistributive protection |
| C7 Transgression-tolerant coercive-order outliers | 5.1% | 59.0% | Unusual tolerance of several transgressions combined with coercive-order responses and low participation |
| C8 Boundary-defensive localists | 4.9% | 51.1% | Strong closure toward racial, religious, linguistic, and immigrant out-groups |
| C9 Institution-trusting civic joiners | 7.7% | 56.4% | High civic and professional membership, agency, information connection, and structural advantage |
| C10 Agentic associational traditionalists | 6.0% | 60.7% | High associational participation and resources combined with religious and personal-choice traditionalism |

The favorable share ranges from 49.1% in C2 to 62.2% in C1, a 13.1-point span. Profile means account for only 0.70% of WVS-weighted individual favorable/unfavorable variance (0.69% with equal-country weighting). No profile is an AI camp: each contains substantial favorable and unfavorable populations, and the rank order does not follow a simple secular-modern versus religious-traditional continuum. C1 is the most favorable, but religious C5, C6, and C10 are close behind; C2 is least favorable while the highly secular C0 sits below the pooled mean.

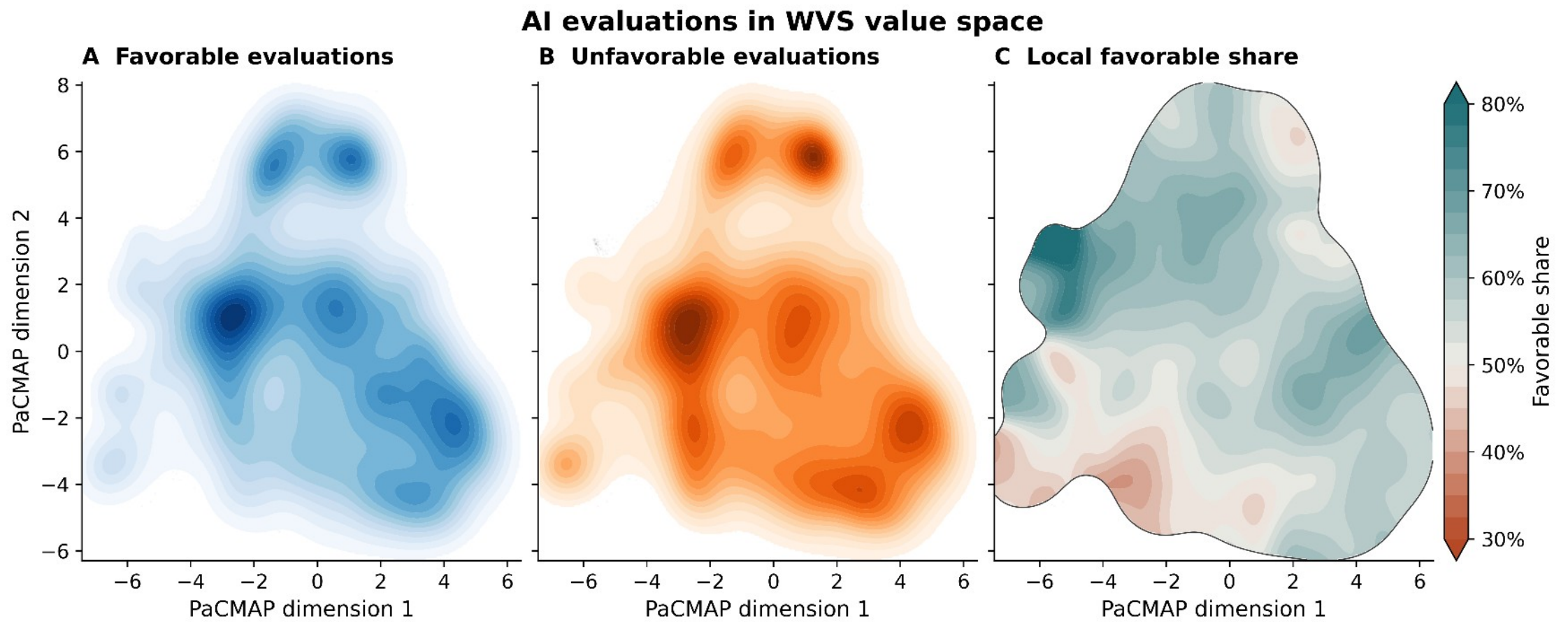

*Figure 4. Overlap between favorable and unfavorable AI evaluations in a two-dimensional PaCMAP representation fitted using the corrected non-AI feature set. Panels A and B show WVS-weighted kernel density estimates, each scaled to its own maximum; darker shading indicates higher density within that class, and color intensity does not indicate relative class prevalence. Panel C shows favorable density divided by the sum of favorable and unfavorable densities, retaining their relative prevalence. The boundary encloses the highest-density region containing 99% of pooled evaluator density; ratios outside it are masked. Both classes use the same Scott bandwidth (0.50 and 0.56 PaCMAP units along dimensions 1 and 2). The display includes 33,803 favorable or unfavorable evaluators, whose overall WVS-weighted favorable share is 57.4%. Other response states are excluded. The substantial spatial overlap is descriptive and was not used for classification or cluster selection. Plotting code was developed with AI assistance.*

### 5.2 Within-profile associations

The main inferential question is how thematic values are associated with AI evaluation within each cross-national respondent profile. Figure 5 estimates one set of profile-specific associations using respondents pooled across countries, with all thematic measures entered jointly. Country normalization defines respondents' value positions relative to their national means, while country intercepts adjust for national differences in baseline evaluation. A row star marks an omnibus difference across profiles after correction; a small dot marks a profile-specific slope that survives the separate cell-level correction.

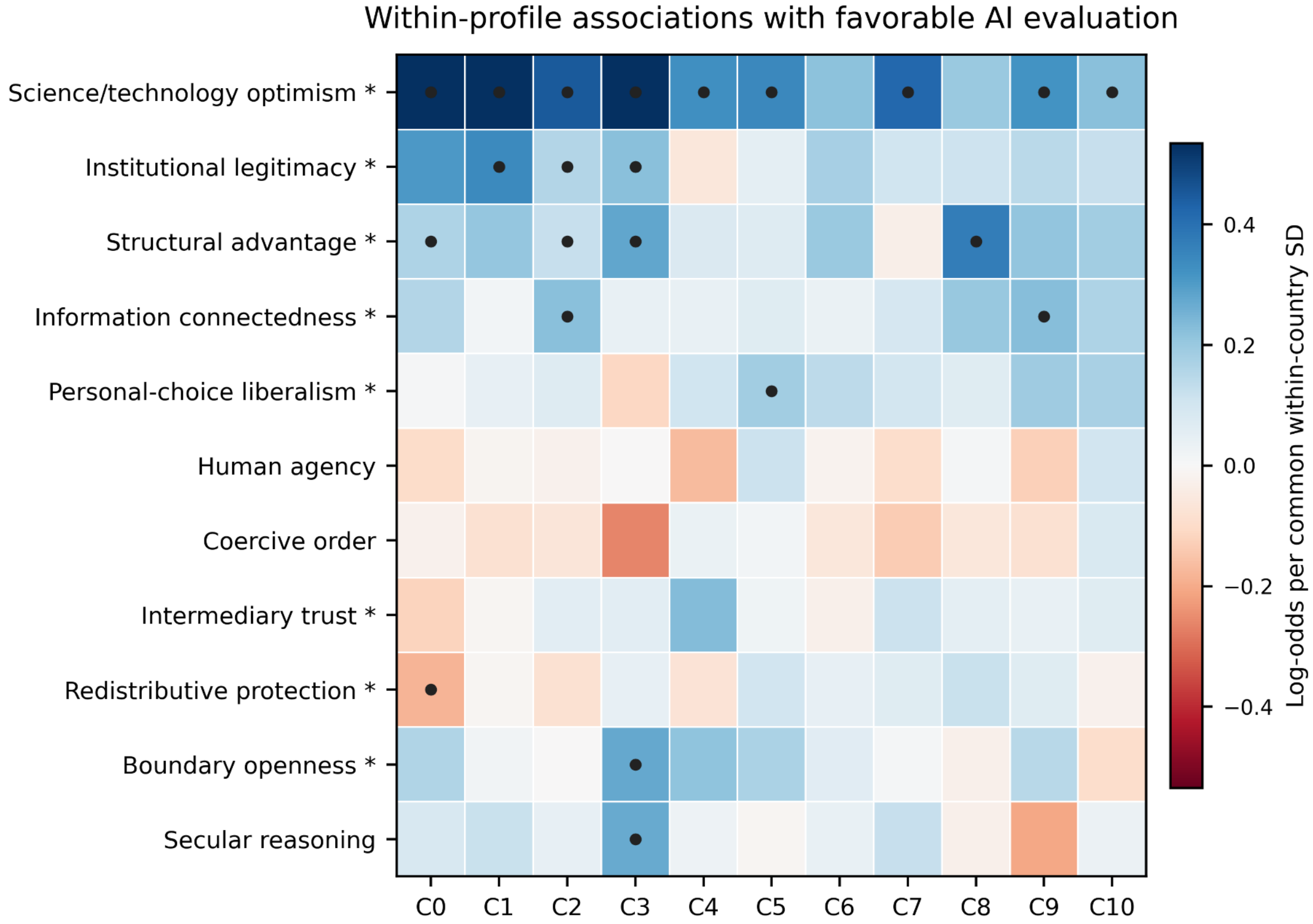

*Figure 5. Within-profile thematic associations with favorable AI evaluation across C0-C10, estimated by pooling respondents across countries. The country-centered thematic scores use one common pooled within-country standard deviation. Profile-specific log-odds coefficients adjust jointly for the other thematic measures and country intercepts. Uncertainty is estimated using 24 country-delete jackknife refits. Row stars denote BH-corrected omnibus profile heterogeneity; small dots denote corrected profile-specific slopes. Plotting code was developed with AI assistance.*

Eight of 11 omnibus tests are significant in the selected partition: science optimism, institutional legitimacy, structural advantage, information connectedness, personal-choice liberalism, intermediary trust, redistributive protection, and boundary openness. That number alone is not the strongest result because the partition was learned from the same respondents. Across three corrected alternative partitions, only science optimism remains omnibus-significant every time. Redistributive protection and secular reasoning recur in two alternatives; every other omnibus result appears in at most one. The publishable conclusion is therefore a robust difference in the strength of the science-optimism association and exploratory evidence of broader profile-conditioned heterogeneity.

Science optimism is positively associated with favorable evaluation in C0, C1, C2, C3, C4, C5, C7, C9, and C10 after cell-level correction. Its average marginal effect ranges from about 4.7 percentage points per common within-country SD in C10 to 16.6 points in C0. It is smaller and imprecise in C6 and C8. A common anchor thus recurs widely while varying substantially in strength.

The other adjusted associations give the profiles more specific texture. In C0, structural advantage is positive and redistributive protection negative. C1 adds institutional legitimacy. C2 combines legitimacy, structural advantage, and information connectedness with science optimism. C3 has the broadest adjusted package: legitimacy, resources, openness, and secular reasoning are all positive. C5 adds personal-choice liberalism. C8 is distinguished primarily by structural advantage, while C9 adds information connectedness. C6 has no coefficient surviving cell-level correction despite its relatively favorable mean. Profile composition and within-profile discrimination are therefore empirically distinct.

The alternative-partition coefficient correlations with the primary solution are .852-.954 and sign agreement is 83.5%-90.9%. A regularized model that also permits country-varying slopes correlates .950 with the primary profile coefficients and agrees on 92.6% of signs. These checks support broad directional continuity while reinforcing caution about claims tied to a particular label or marginal coefficient.

The all-country-component sensitivity is closer to the primary fit: coefficient correlation is .994, sign agreement is 95.9%, and science optimism remains among five corrected omnibus axes. In the 16-country 2024-start sensitivity, correlation is .935 and sign agreement is 86.8%; only science optimism and personal-choice liberalism retain corrected omnibus differences. Science optimism is therefore the only profile-heterogeneity result that survives every partition, item-harmonization, and fieldwork-period check.

The two equal-country profile reconstructions span 50.5%-67.6% favorable evaluation and place only .81%-1.01% of individual evaluation variance between their means. Although those values are slightly above the primary .70%, they preserve the central weak-sorting result despite substantially different memberships.

# 6. Country Value Contexts

## 6.1 Three outcome-independent contexts

The country hierarchy asks whether samples with similar respondent-profile mixtures and mean thematic positions form broader comparative contexts. Table 4 reports the three groups and their response patterns. The selected K=3 solution has silhouette .300; respondent-bootstrap ARI averages .975, PAM ARI is .780, the profile-only solution exactly reproduces the fused partition, and the axis-only solution has ARI .733. A K=4 refinement is available in the research materials but is not needed for the main analysis.

*Table 4. Outcome-independent groups of countries and response patterns.*

| Context | Countries | Narrow evaluability | Favorable among evaluators |
|---|---|---|---|
| R1 Religious-traditional | BGD, EGY, IDN, MAR | 82.5% | 65.7% |
| R2 Mixed traditional-organizational | BOL, CIV, COL, GTM, KEN, MYS, NGA, PAK, PER, ZAF | 87.0% | 57.3% |
| R3 Secular-emancipative | AND, CAN, GRC, JPN, SGP, SVK, SWE, THA, TWN, USA | 94.2% | 61.0% |

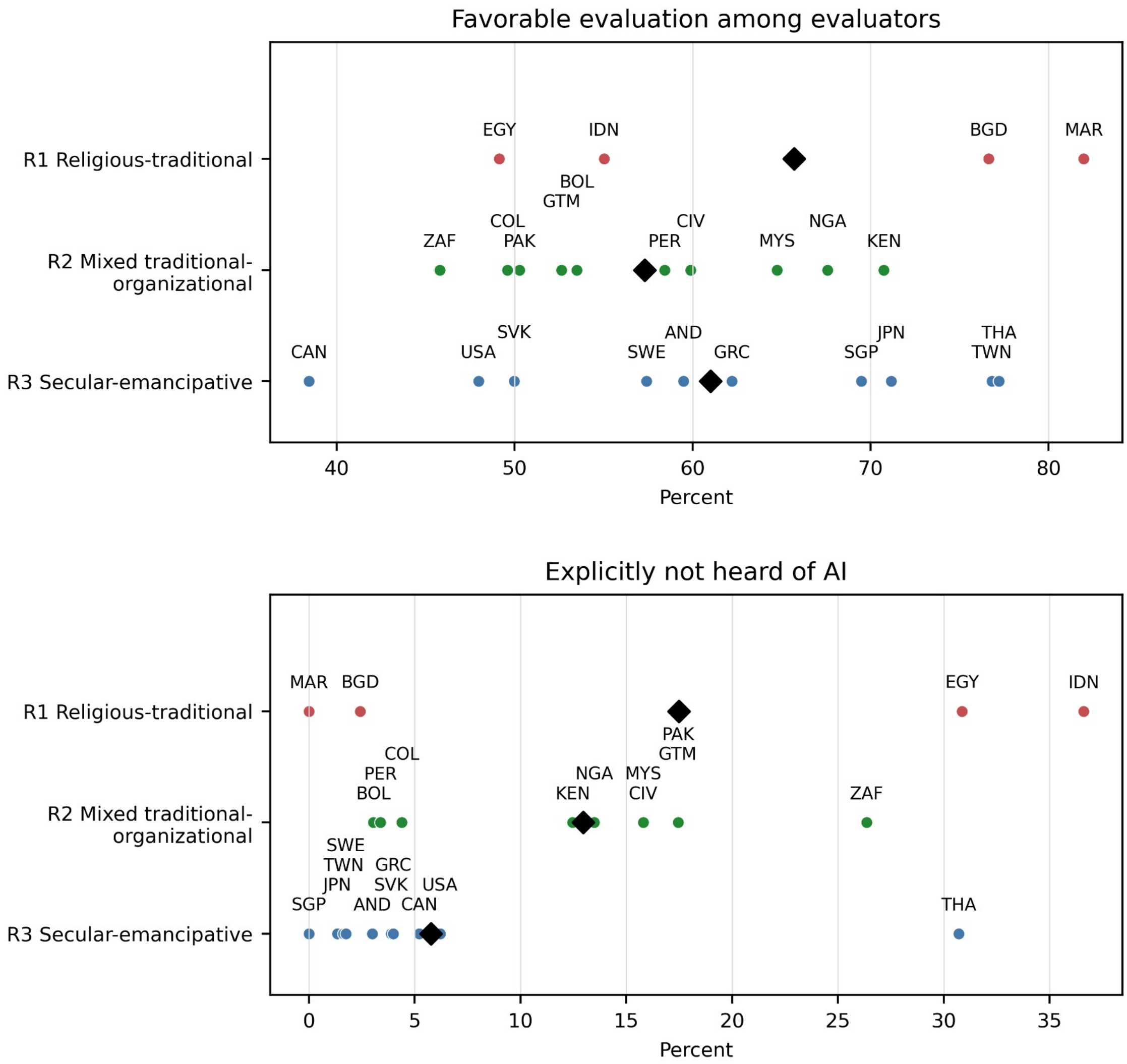

*Figure 6A. Country estimates of favorable evaluation among evaluators and the share of all respondents explicitly coded not heard of AI, grouped by outcome-independent context. Black diamonds weight countries equally. The not-heard response is narrower than a general awareness scale and is not interchangeable with don't know or nonresponse. Plotting code was developed with AI assistance.*

R1 is dominated by C6 religious restrictive material solidarists and is low on science optimism, structural advantage, information connectedness, personal choice, and secular reasoning, while high on redistributive protection. R2 has the most mixed profile composition, with C5, C6, C9, and C10 relatively prevalent; it is higher on coercive order and lower on institutional legitimacy, secular reasoning, and redistribution. R3 concentrates C0, C1, and C3 and is higher on secular reasoning, personal choice, structural advantage, science optimism, and boundary openness while lower on coercive order.

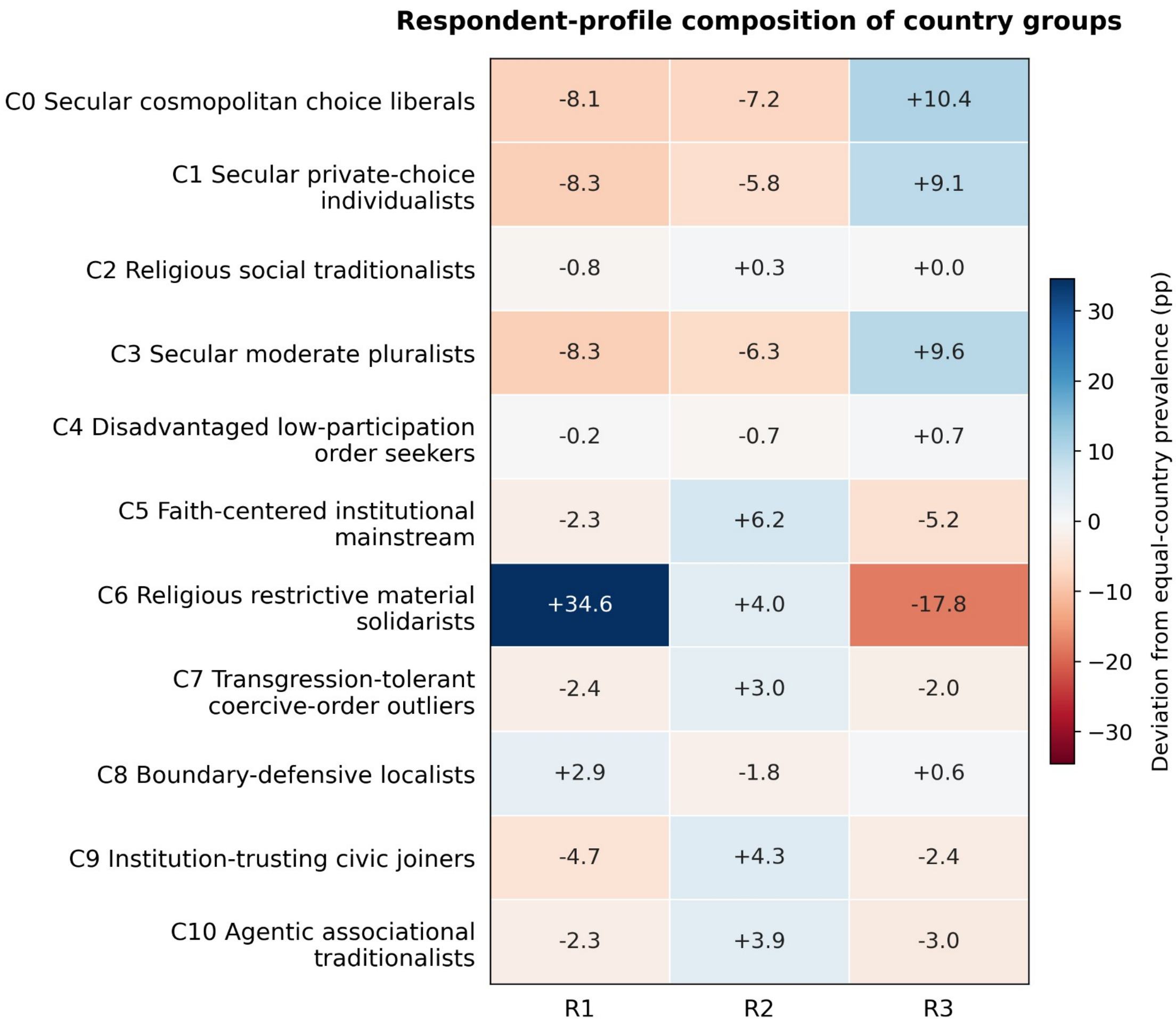

*Figure 6B. Respondent-profile composition of the three outcome-independent groups of countries. Cells show departures in profile shares from their prevalence across all 24 countries, weighting countries equally, in percentage points. Blue denotes greater prevalence; red denotes lower prevalence. Plotting code was developed with AI assistance.*

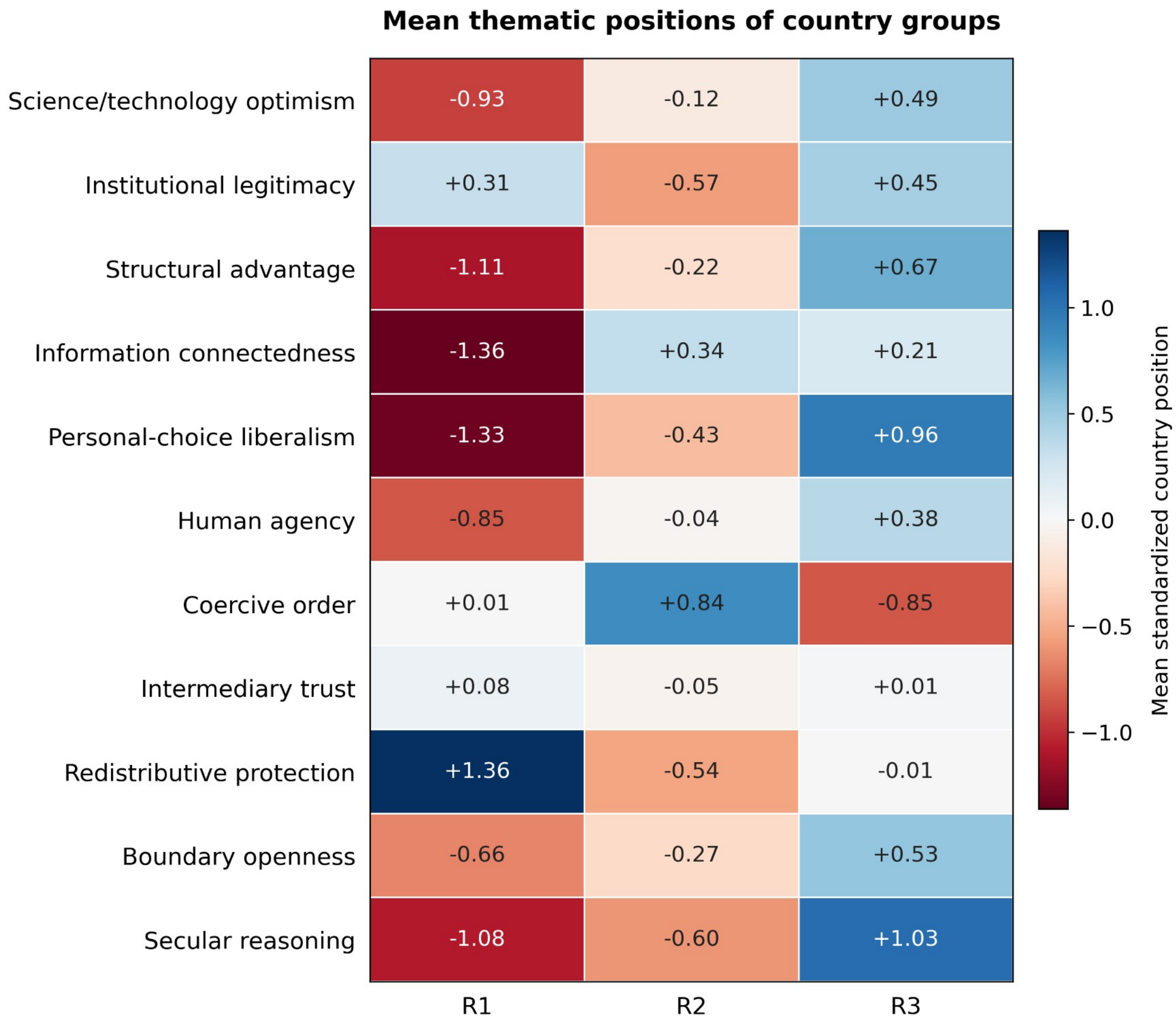

*Figure 6C. Mean thematic positions of the three outcome-independent groups of countries. Cells average standardized country means, weighting countries equally. Blue denotes a higher position; red denotes a lower position. These ecological descriptions, like the profile shares in Figure 6B, are not individual-level slopes. Plotting code was developed with AI assistance.*

The three groups are not AI-evaluation tiers. R1 has the highest context mean despite the lowest science optimism and information connection, and substantial country differences remain within every context. Its four-country favorable mean also conditions on a more selected evaluator population, particularly in Indonesia and Egypt. This is precisely why context composition and within-context associations must be separated.

### 6.2 Context-specific associations

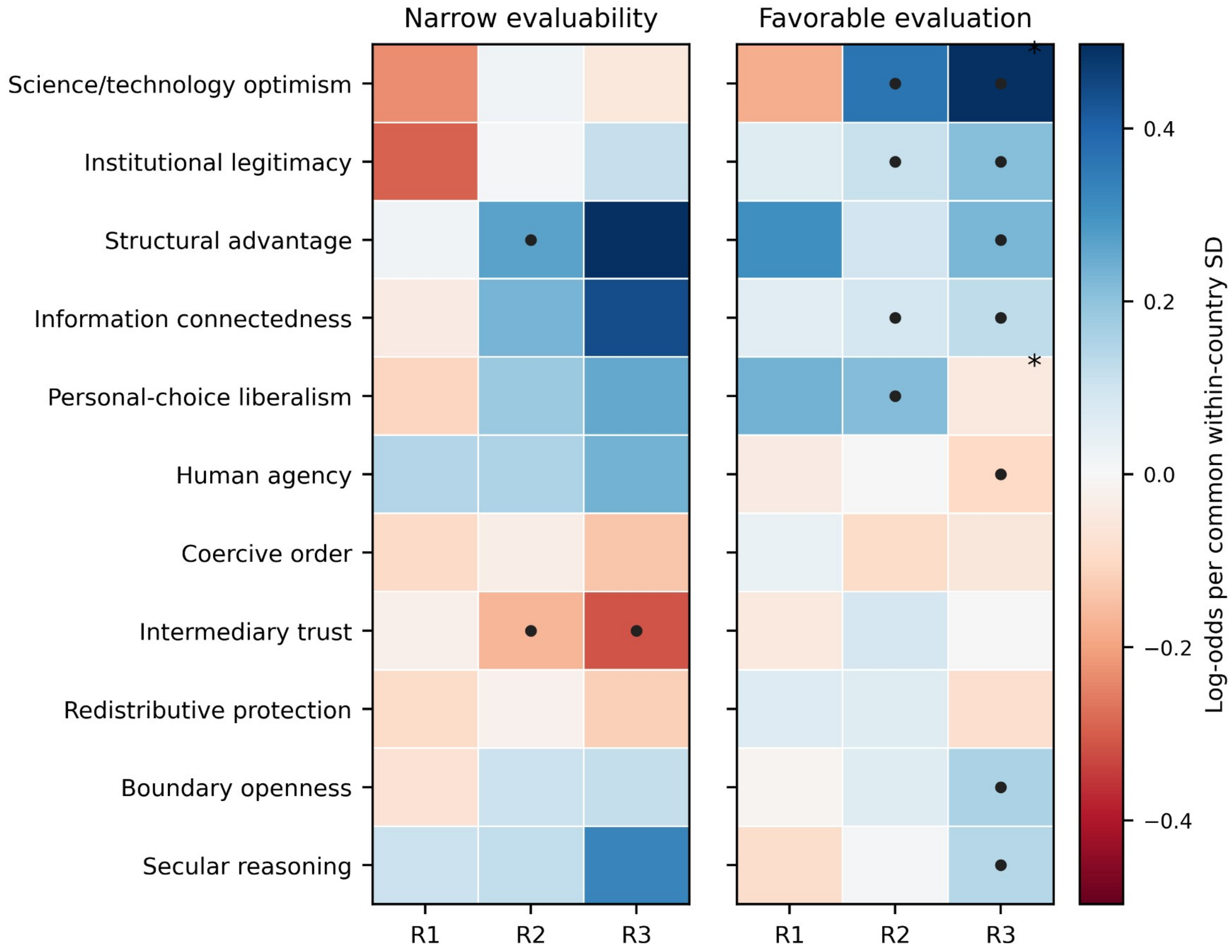

*Figure 7. Joint within-country thematic associations with narrow evaluability and favorable evaluation across R1-R3. Coefficients use a common pooled within-country unit and country fixed intercepts. Row stars mark BH-corrected omnibus context heterogeneity; small dots mark corrected context-specific slopes. Plotting code was developed with AI assistance.*

No evaluability axis has a significant omnibus context interaction after correction. Although structural advantage is positive in R2 and intermediary trust is negative in R2 and R3 at the cell level, the evidence does not establish that those slopes differ across contexts. This sharply limits the earlier stage-based narrative.

For favorable evaluation, only science optimism and personal-choice liberalism have corrected omnibus context differences. Science optimism is negative but imprecise in R1 (`b=-.179`), positive in R2 (`b=.364`; average marginal effect 8.0 points per common SD), and strongest in R3 (`b=.748`; 14.3 points). Personal-choice liberalism is positive in R1 and R2 but near zero and negative in R3; only R2's cell estimate survives correction. Secular reasoning narrowly misses the omnibus threshold (`q=.052`) despite a negative-to-positive descriptive progression from R1 to R3.

Within R2, favorable evaluation is also positively associated with institutional legitimacy and information connectedness after cell-level correction. R3 contains the broadest adjusted package: legitimacy, structural advantage, information connectedness, boundary openness, and secular reasoning are positive, while human agency is modestly negative. These cell findings describe each context but should not be mistaken for significant cross-context differences unless the omnibus test also supports that claim.

The harmonized-item context coefficients correlate .990 with the primary estimates and agree on 92.4% of signs; science optimism and personal choice remain heterogeneous, and secular reasoning additionally crosses the corrected threshold. In the 16-country 2024-start sensitivity, coefficient correlation is .873 and sign agreement is 87.9%; only science optimism remains omnibus-significant. The science result is robust, while the personal-choice context difference is less stable to fieldwork-period composition.

# 7. Discussion

## 7.1 Science optimism as a recurrent cross-country anchor

The clearest substantive regularity is that a general belief in the benefits of science and technology travels widely into AI evaluation. Q118 dominates pooled prediction, its removal costs about .020-.022 AUC even after duplicate indices are removed, and it recurs in national top-ten lists far more consistently than any other source. This supports prior work linking support for science, innovation worldviews, and perceived technological benefit to AI attitudes (O'Shaughnessy et al. 2023; Veckalov et al. 2023; Rice and Wu 2025).

Q118 and Q121 capture related but distinct judgments. AI can be understood as qualitatively different from earlier technologies because it raises specific concerns about autonomous action, opacity, surveillance, employment, concentrated power, and delegated decisions. The profile and country-context models show that broader science optimism carries into AI evaluation across most settings while varying meaningfully in strength. The smaller adjusted associations in C6 and C8 and the negative, imprecise estimate in R1 locate configurations in which favorable AI judgment draws more heavily on orientations beyond generalized technological progress.

## 7.2 Secondary regularities and national combinations

The secondary pooled findings identify several substantive patterns around the science-optimism anchor. Receptivity to expert decision making connects favorable evaluation to comfort with delegated technical authority, consistent with research on algorithm appreciation and AI governance (Logg, Minson, and Moore 2019; David, Choung, and Seberger 2024). Education, email use, and information connection point to a capacity and incorporation gradient. Immigration openness, universal contraceptive access, and boundary openness connect favorable evaluation to selected liberal and cosmopolitan orientations. Interpersonal and business confidence add a trust component.

These orientations combine differently across settings. Confidence in government or political leaders need not resemble business confidence; personal-choice liberalism is consequential in selected profiles and contexts; human agency is slightly negative in R3 after adjacent values enter jointly. These results extend work on trust and cultural values by showing how sources and surrounding constructs acquire distinct roles (Chen and Wen 2021; Schepman and Rodway 2023). The resulting picture is a differentiated set of recurring associations rather than a single social constituency.

## 7.3 Values structure evaluation within social profiles

Person-centered AI research often uses AI-specific items to form enthusiast, skeptic, or ambivalent profiles (Shum and Lau 2024). The present analysis instead forms profiles from broader values and social position, allowing AI evaluation to be examined as an outcome. Their 13.1-point range establishes modest differences in average evaluation but the 99.3% of individual variance within profiles directs attention to the determinants of each profile's internal AI divide.

This pattern accords with theories of weak belief-system constraint (Converse 1964; Elder and O'Brian 2022). AI touches science, work, surveillance, administration, development, morality, and agency, but those implications need not be joined by the same political cues or group alignments in every national sample. The profiles are useful because they distinguish **defining values** from **discriminating values**. Religiosity or secularism may define a profile without explaining its internal AI divide; resources or institutional legitimacy may discriminate evaluation without being the profile's most distinctive trait.

## 7.4 Multiple routes through governed modernity

The country contexts qualify narratives presenting a single modernization ladder. R1 combines religious restriction and lower resources with the highest favorable mean; R3 combines secular-emancipative values and greater resources with the strongest science-optimism slope; R2 occupies neither a simple midpoint nor a residual category. Scientific development, authority, social protection, institutional capacity, and individual choice can be recombined in distinct projects of social change, consistent with multiple-modernities arguments (Eisenstadt 2000).

### 7.5 Implications for anticipating AI receptivity and governance

The findings have three implications for anticipating the social reception of AI. First, national averages are insufficient guides to public receptivity: the same country contains multiple value configurations, and most favorable/unfavorable variation lies within rather than between those profiles. Second, general optimism about science and technology is the most dependable cross-country signal, but its association with AI evaluation varies enough that progress-oriented appeals cannot be assumed to work uniformly. Third, institutional legitimacy, resources, information connection, authority, and openness matter in different combinations. Organizations and policymakers introducing or governing AI should therefore diagnose the expectations surrounding a particular setting rather than infer receptivity from a generic national culture score or a single trust measure.

### 7.6 Research directions: how value profiles interpret AI

The within-profile associations open a research agenda on how people connect AI to preferred forms of social order. Governed modernity suggests that values may relate to AI evaluation partly through whether AI is imagined to reinforce, reform, or threaten valued institutions and practices. This extends the link between technological futures and collective purposes in sociotechnical-imaginaries research (Jasanoff and Kim 2015), treating the profiles as starting points for investigation rather than evidence of shared imaginaries.

First, institutional legitimacy motivates a distinction between stewardship and substitution. Confidence in institutions could accompany favorable evaluation when AI is understood as an instrument of competent governance; distrust could encourage receptivity when AI is imagined as replacing biased or ineffective human authorities. Molina and Sundar’s (2024) content-moderation findings establish the plausibility of greater receptivity to machines among people who distrust humans, but do not establish this institutional mechanism. A testable proposition is that the legitimacy association depends on whether AI assists trusted authorities or substitutes for distrusted decision-makers.

Second, the secular-reasoning contrast between C3 secular moderate pluralists and C9 institution-trusting civic joiners raises a question about emancipation and delegation. The positive C3 association survives correction; C9’s negative estimate is suggestive. One hypothesis is that secular reasoning aligns with favorable evaluation when AI represents distance from inherited authority, but with concern when AI represents delegation away from accountable human institutions. Differences in exposure or the applications respondents imagine offer competing explanations that deeper survey work could distinguish.

Third, redistributive protection and personal-choice associations invite tests of whether AI is understood as collective advancement or concentrated power. Expected gains in public services, opportunity, and individual control may connect these values to favorable judgment, whereas expected concentration of ownership, weakened social protection, or surveillance may connect them to concern. This proposition makes ownership, benefit distribution, and human oversight substantive conditions of receptivity rather than incidental details of deployment.

Research could begin with interviews and open-ended surveys of favorable and unfavorable evaluators in contrasting value configurations within the same countries. An independent survey experiment could then hold technical capabilities constant while varying the deploying institution, ownership, benefit distribution, and human oversight. Preregistered configuration-by-treatment comparisons would test the proposed relationships without assuming that the exact eleven-profile partition is universal. Measuring respondents’ interpretations would connect the comparative map to specific hypotheses about how AI deployment acquires social legitimacy.

## 8. Limitations

The cross-sectional design identifies associations, not causal effects or mediation. The single AI item captures broad societal evaluation rather than adoption, regulation, or application-specific judgments, and evaluation models condition on favorable or unfavorable responses. Cross-language equivalence remains untested; the thematic measures are formative summaries with uneven internal coherence.

The 24 provisional country samples are not globally representative, and equal-country weighting targets their average rather than a global population. Country-delete uncertainty does not reconstruct national survey designs, and fieldwork timing is confounded with country composition. Correlated predictors can exchange SHAP attribution, limiting the interpretation of detailed national rankings.

Respondent profiles and country groups are data-adaptive descriptions, not fixed population types. Profile memberships and some secondary associations vary with representation, country balance, and fieldwork restrictions; inference is conditional on the selected partition. The analysis is exploratory rather than preregistered. Sensitivity checks support the recurrence of science optimism and limited between-profile sorting more consistently than individual secondary contrasts; detailed diagnostics are reported in the appendix.

## 9. Conclusion

Across 24 provisional WVS Wave 8 country samples, a broad evaluation of whether AI is good or bad for humanity has a recognizable cross-country architecture. General optimism about science and technology is its strongest and most recurrent predictive anchor. Expert authority, education and information access, openness, selected forms of trust, and demographic position add meaningful but less stable regularities.

The eleven profiles show how those correlates are recombined across recognizable social configurations. Their compositions and mean evaluations differ, while the concentration of favorable/unfavorable variation within profiles makes profile-specific determinants substantively central. Science optimism supplies the most consistent source of heterogeneity across profile sensitivities, and personal-choice liberalism joins it in distinguishing the three country contexts.

Generalized AI opinion is organized by common anchors operating through different social configurations. Their strength and surrounding associations depend on how scientific progress, institutional competence, resources, authority, openness, and human agency are combined. For the study of technological change, this means that public receptivity cannot be read directly from national averages or a universal adoption model. Comparative research can instead recover both the regularities that travel and the context-specific pathways through which emerging technologies acquire social legitimacy.

## Declarations

**Originality and author approval.** This manuscript is original and has not been published elsewhere. Neither this manuscript nor any variant is under review at another journal. All authors have approved this manuscript.

**Data availability.** The analysis uses provisional WVS Wave 8 data available as of April 2026, source `WVS8_v1_Spss_eng_v20260421`. Requests for access to this exact provisional version should be directed to the World Values Survey Association through its data archive. WVS conditions of use require non-profit use and appropriate citation and prohibit redistribution of the data files. The research-materials supplement therefore excludes respondent-level data; the appendix identifies the source and construction rules.

**Code availability.** Analysis code, operational metadata, variable-construction materials, aggregate results, figures, environment specifications, and integrity manifests are archived in Zenodo, version 1.0 (Flanders and Nungsari 2026), at https://doi.org/10.5281/zenodo.23688866. Original code is licensed under MIT; original documentation and outputs are licensed under CC BY 4.0. Attributed WVS metadata and other third-party content are excluded from these grants. Respondent-level data and individual-record derivatives are not redistributed. Replication requires authorized access to the specified WVS source; no generative-AI API calls are needed for the quantitative analyses.

**Ethics.** This study uses secondary, de-identified World Values Survey data and involved no new participant recruitment or data collection. Institutional review board review is not required under the authors' institutional policy for secondary analysis of existing sources such as the WVS.

**Funding.** This research did not receive any specific grant from funding agencies in the public, commercial, or not-for-profit sectors.

**Competing interests.** The authors declare no competing interests.

## Declaration of Generative AI and AI-Assisted Technologies

The authors used OpenAI Codex and ChatGPT for substantial assistance with methodological ideation, development and debugging of analysis and visualization code, and substantive manuscript drafting and revision. GLM-5.2 (`z-ai/glm-5.2`) was also used for exploratory API-based summaries and interpretations, as described in Section 3.8. The authors directed the research, reviewed and revised AI-assisted material, and take responsibility for the methods, results, interpretations, and manuscript. AI-generated text is not treated as independent empirical evidence.

## References

Bakk, Z., and J. Kuha. 2021. "Relating Latent Class Membership to External Variables: An Overview." *British Journal of Mathematical and Statistical Psychology* 74(2): 340-362. https://doi.org/10.1111/bmsp.12227.

Bell, A., and K. Jones. 2015. "Explaining Fixed Effects: Random Effects Modeling of Time-Series Cross-Sectional and Panel Data." *Political Science Research and Methods* 3(1): 133-153. https://doi.org/10.1017/psrm.2014.7.

Benjamini, Y., and Y. Hochberg. 1995. "Controlling the False Discovery Rate: A Practical and Powerful Approach to Multiple Testing." *Journal of the Royal Statistical Society: Series B* 57(1): 289-300. https://doi.org/10.1111/j.2517-6161.1995.tb02031.x.

Bergman, L. R., and K. Trost. 2006. "The Person-Oriented versus the Variable-Oriented Approach: Are They Complementary, Opposites, or Exploring Different Worlds?" *Merrill-Palmer Quarterly* 52(3): 601-632. https://doi.org/10.1353/mpq.2006.0023.

Busuioc, M., and M. Maggetti. 2026. "Worthy of Trust? AI Governance and the Role of (Dis)Trust." *Perspectives on Public Management and Governance*: gvag007. https://doi.org/10.1093/ppmgov/gvag007.

Castelo, N., M. W. Bos, and D. R. Lehmann. 2019. "Task-Dependent Algorithm Aversion." *Journal of Marketing Research* 56(5): 809-825. https://doi.org/10.1177/0022243719851788.

Chen, M.-F., and X. Wen. 2021. "Impacts of Attitudes Toward Government and Corporations on Public Trust in Artificial Intelligence." *Communication Studies* 72(1): 115-131. https://doi.org/10.1080/10510974.2020.1807380.

Chen, T., and C. Guestrin. 2016. "XGBoost: A Scalable Tree Boosting System." In *Proceedings of the 22nd ACM SIGKDD International Conference on Knowledge Discovery and Data Mining*, 785-794. https://doi.org/10.1145/2939672.2939785.

Choung, H., P. David, and A. Ross. 2023. "Trust and Ethics in AI." *AI & Society* 38(2): 733-745. https://doi.org/10.1007/s00146-022-01473-4.

Choung, H., P. David, and J. S. Seberger. 2024. "A Multilevel Framework for AI Governance." In *The Routledge Handbook of Global and Digital Governance Crossroads: Stakeholder Engagement and Democratization*, edited by P. Shroff-Mehta, J. N. Sheth, J. Garrison, and S. R. Mehta, 310-323. Routledge. https://doi.org/10.4324/9781003316077-25.

Converse, P. E. 1964. "The Nature of Belief Systems in Mass Publics." In *Ideology and Discontent*, edited by D. E. Apter, 206-261. New York: Free Press.

David, P., H. Choung, and J. S. Seberger. 2024. "Who Is Responsible? US Public Perceptions of AI Governance through the Lenses of Trust and Ethics." *Public Understanding of Science* 33(5): 654-672. https://doi.org/10.1177/09636625231224592.

Dietvorst, B. J., J. P. Simmons, and C. Massey. 2015. "Algorithm Aversion: People Erroneously Avoid Algorithms after Seeing Them Err." *Journal of Experimental Psychology: General* 144(1): 114-126. https://doi.org/10.1037/xge0000033.

Eisenstadt, S. N. 2000. "Multiple Modernities." *Daedalus* 129(1): 1-29.

Enders, C. K., and D. Tofighi. 2007. "Centering Predictor Variables in Cross-Sectional Multilevel Models: A New Look at an Old Issue." *Psychological Methods* 12(2): 121-138. https://doi.org/10.1037/1082-989X.12.2.121.

Elder, E. M., and N. A. O'Brian. 2022. "Social Groups as the Source of Political Belief Systems: Fresh Evidence on an Old Theory." *American Political Science Review* 116(4): 1407-1424. https://doi.org/10.1017/S0003055422000326.

Flanders, S., and M. Nungsari. 2026. *Research Materials for Conditional Values: The Contingent Foundations of Attitudes toward Artificial Intelligence*. Version 1.0. Research materials. Zenodo. https://doi.org/10.5281/zenodo.23688866.

Georgieff, A. 2024. "Artificial Intelligence and Wage Inequality." *OECD Artificial Intelligence Papers* No. 13. Paris: OECD Publishing. https://doi.org/10.1787/bf98a45c-en.

Georgieff, A., and R. Hyee. 2021. "Artificial Intelligence and Employment: New Cross-Country Evidence." *OECD Social, Employment and Migration Working Papers* No. 265. Paris: OECD Publishing. https://doi.org/10.1787/c2c1d276-en.

Gillespie, N., S. Lockey, T. Ward, A. Macdade, and G. Hassed. 2025. *Trust, Attitudes and Use of Artificial Intelligence: A Global Study 2025.* University of Melbourne and KPMG. https://mbs.edu/-/media/PDF/Research/Trust_in_AI_Report.pdf.

Heisig, J. P., M. Schaeffer, and J. Giesecke. 2017. "The Costs of Simplicity: Why Multilevel Models May Benefit from Accounting for Cross-Cluster Differences in the Effects of Controls." *American Sociological Review* 82(4): 796-827. https://doi.org/10.1177/0003122417717901.

Inglehart, R. 1997. *Modernization and Postmodernization: Cultural, Economic, and Political Change in 43 Societies.* Princeton, NJ: Princeton University Press.

Inglehart, R., and C. Welzel. 2005. *Modernization, Cultural Change, and Democracy: The Human Development Sequence.* Cambridge: Cambridge University Press.

Jasanoff, S., and S.-H. Kim, eds. 2015. *Dreamscapes of Modernity: Sociotechnical Imaginaries and the Fabrication of Power.* Chicago: University of Chicago Press.

Kelly, S., S.-A. Kaye, and O. Oviedo-Trespalacios. 2023. "What Factors Contribute to the Acceptance of Artificial Intelligence? A Systematic Review." *Telematics and Informatics* 77: 101925. https://doi.org/10.1016/j.tele.2022.101925.

Koenig, P. D. 2025. "Attitudes toward Artificial Intelligence: Combining Three Theoretical Perspectives on Technology Acceptance." *AI & Society* 40(3): 1333-1345. https://doi.org/10.1007/s00146-024-01987-z.

Lahusen, C., M. Maggetti, and M. Slavkovik. 2024. "Trust, Trustworthiness and AI Governance." *Scientific Reports* 14: 20752. https://doi.org/10.1038/s41598-024-71761-0.

Lane, M., and A. Saint-Martin. 2021. "The Impact of Artificial Intelligence on the Labour Market: What Do We Know So Far?" *OECD Social, Employment and Migration Working Papers* No. 256. Paris: OECD Publishing. https://doi.org/10.1787/7c895724-en.

Logg, J. M., J. A. Minson, and D. A. Moore. 2019. "Algorithm Appreciation: People Prefer Algorithmic to Human Judgment." *Organizational Behavior and Human Decision Processes* 151: 90-103. https://doi.org/10.1016/j.obhdp.2018.12.005.

Longoni, C., A. Bonezzi, and C. K. Morewedge. 2019. "Resistance to Medical Artificial Intelligence." *Journal of Consumer Research* 46(4): 629-650. https://doi.org/10.1093/jcr/ucz013.

Lundberg, S. M., and S.-I. Lee. 2017. "A Unified Approach to Interpreting Model Predictions." In *Advances in Neural Information Processing Systems* 30. https://proceedings.neurips.cc/paper_files/paper/2017/hash/8a20a8621978632d76c43dfd28b67767-Abstract.html.

MacKinnon, J. G., M. O. Nielsen, and M. D. Webb. 2025. "Cluster-Robust Jackknife and Bootstrap Inference for Logistic Regression Models." *Econometric Reviews*. https://doi.org/10.1080/07474938.2025.2515161.

Molina, M. D., and S. S. Sundar. 2024. "Does Distrust in Humans Predict Greater Trust in AI? Role of Individual Differences in User Responses to Content Moderation." *New Media & Society* 26(6): 3638-3656. https://doi.org/10.1177/14614448221103534.

Mize, T. D. 2019. "Best Practices for Estimating, Interpreting, and Presenting Nonlinear Interaction Effects." *Sociological Science* 6: 81-117. https://doi.org/10.15195/v6.a4.

Mongrain, P., and S. Walgrave. 2026. "One-Dimensional, Multidimensional, or Non-Dimensional? Ideological Structure in Mass and Elite Opinion." *European Journal of Political Research*, advance online publication. https://doi.org/10.1017/S1475676526100838.

Nussberger, A.-M., L. Luo, L. E. Celis, and M. J. Crockett. 2022. "Public Attitudes Value Interpretability but Prioritize Accuracy in Artificial Intelligence." *Nature Communications* 13: 5821. https://doi.org/10.1038/s41467-022-33417-3.

Nylund-Gibson, K., R. P. Grimm, and K. E. Masyn. 2019. "Prediction from Latent Classes: A Demonstration of Different Approaches to Include Distal Outcomes in Mixture Models." *Structural Equation Modeling: A Multidisciplinary Journal* 26(6): 967-985. https://doi.org/10.1080/10705511.2019.1590146.

O'Shaughnessy, M. R., D. S. Schiff, L. R. Varshney, C. J. Rozell, and M. A. Davenport. 2023. "What Governs Attitudes toward Artificial Intelligence Adoption and Governance?" *Science and Public Policy* 50(2): 161-176. https://doi.org/10.1093/scipol/scac056.

Rice, R. E., and M.-Y. Wu. 2025. "Difference in and Influences on Public Opinion about Artificial Intelligence in 20 Economies: Reducing Uncertainty through Awareness, Knowledge, and Trust." *International Journal of Communication* 19: 740-765. https://ijoc.org/index.php/ijoc/article/view/22909.

Sartori, L., and G. Bocca. 2023. "Minding the Gap(s): Public Perceptions of AI and Socio-Technical Imaginaries." *AI & Society* 38(2): 443-458. https://doi.org/10.1007/s00146-022-01422-1.

Schepman, A., and P. Rodway. 2020. “Initial Validation of the General Attitudes towards Artificial Intelligence Scale.” *Computers in Human Behavior Reports* 1: 100014. https://doi.org/10.1016/j.chbr.2020.100014.

Schepman, A., and P. Rodway. 2023. “The General Attitudes towards Artificial Intelligence Scale (GAAIS): Confirmatory Validation and Associations with Personality, Corporate Distrust, and General Trust.” *International Journal of Human-Computer Interaction* 39(13): 2724-2741. https://doi.org/10.1080/10447318.2022.2085400.

Shum, N.-Y. E., and H.-P. B. Lau. 2024. “Perils, Power and Promises: Latent Profile Analysis on the Attitudes towards Artificial Intelligence (AI) among Middle-Aged and Older Adults in Hong Kong.” *Computers in Human Behavior: Artificial Humans* 2(2): 100091. https://doi.org/10.1016/j.chbah.2024.100091.

Stein, J.-P., T. Messingschlager, T. Gnambs, F. Hutmacher, and M. Appel. 2024. “Attitudes towards AI: Measurement and Associations with Personality.” *Scientific Reports* 14: 2909. https://doi.org/10.1038/s41598-024-53335-2.

Veckalov, B., A. van Stekelenburg, F. van Harreveld, and B. T. Rutjens. 2023. “Who Is Skeptical About Scientific Innovation? Examining Worldview Predictors of Artificial Intelligence, Nanotechnology, and Human Gene Editing Attitudes.” *Science Communication* 45(3): 337-366. https://doi.org/10.1177/10755470231184203.

Vesely, S., and B. Kim. 2024. “Survey Evidence on Public Support for AI Safety Oversight.” *Scientific Reports* 14: 31491. https://doi.org/10.1038/s41598-024-82977-5.

Wang, H., B. Zhao, Y. Jiao, C. Liang, and G. Zheng. 2026. “The Moderating Effect of National Culture on the Relationships between Antecedents and Attitude and Intention for AI Adoption: A Meta-Analysis.” *Technological Forecasting and Social Change* 231: 124803. https://doi.org/10.1016/j.techfore.2026.124803.

Wang, S. 2025. “Public Perceptions of Artificial Intelligence in 20 Countries: Assessing Individual- and Country-Level Factors.” *Cross-Cultural Research* 59(5): 651-676. https://doi.org/10.1177/10693971251336803.

Wang, Y., H. Huang, C. Rudin, and Y. Shaposhnik. 2021. “Understanding How Dimension Reduction Tools Work: An Empirical Approach to Deciphering t-SNE, UMAP, TriMap, and PaCMAP for Data Visualization.” *Journal of Machine Learning Research* 22(201): 1-73. https://jmlr.org/papers/v22/20-1061.html.

Ward, J. H., Jr. 1963. “Hierarchical Grouping to Optimize an Objective Function.” *Journal of the American Statistical Association* 58(301): 236-244. https://doi.org/10.1080/01621459.1963.10500845.

Woo, S. E., J. Hofmans, B. Wille, and L. Tay. 2024. “Person-Centered Modeling: Techniques for Studying Associations between People Rather than Variables.” *Annual Review of Organizational Psychology and Organizational Behavior* 11(1): 453-480. https://doi.org/10.1146/annurev-orgpsych-110721-045646.

World Values Survey Association. 2026. *World Values Survey Wave 8: Provisional Cross-National Data Available as of April 2026.* Source release `WVS8_v1_Spss_eng_v20260421`. https://www.worldvaluessurvey.org/WVSContents.jsp?CMSID=Documentation.

# Supplementary Appendix: Data, Methods, and Extended Results

Abbreviations: AI, artificial intelligence; AUC, area under the receiver-operating-characteristic curve; SHAP, SHapley Additive exPlanations; LOCO, leave-one-country-out; PaCMAP, Pairwise Controlled Manifold Approximation Projection; ARI, adjusted Rand index; PAM, partitioning around medoids; BH, Benjamini-Hochberg false-discovery-rate correction; SD, standard deviation. Profile labels C0-C10 and country-context labels R1-R3 match the manuscript.

## A1. Source, outcome, and response-state audit

The analysis uses provisional World Values Survey (WVS) Wave 8 data available as of April 2026, source `WVS8_v1_Spss_eng_v20260421`. The 24 samples contain 40,960 respondents. Access requests for this exact provisional version should be directed to the WVS data archive. Its conditions of use require non-profit use and appropriate citation and prohibit redistribution of data files. No respondent-level records are included in the research-materials supplement.

| Q121 state | N | Raw share | WVS-weighted share |
|---|---|---|---|
| Explicitly not heard of AI | 4,730 | 11.55% | 11.80% |
| Unfavorable evaluation | 14,414 | 35.19% | 35.30% |
| Favorable evaluation | 19,389 | 47.34% | 46.93% |
| Don't know | 1,963 | 4.79% | 4.82% |
| No answer | 392 | 0.96% | 0.98% |
| Not asked or unavailable | 72 | 0.18% | 0.18% |

Primary models compare favorable with unfavorable evaluation. The secondary first-stage outcome is **narrow evaluability**: Q121 in {1,2} versus Q121=0 among respondents in {0,1,2}. It is not a general awareness measure. Seven samples have at least 10% indeterminate/nonresponse states; Morocco and Singapore contain no explicit not-heard responses. The largest change when don't know is treated as nonevaluation is 32.4 percentage points.

Artifacts:

- `q121_country_comparability/q121_country_response_audit.csv`
- `q121_country_comparability/q121_response_state_totals.csv`
- `q121_country_comparability/q121_response_states_by_country.png`
- `q121_country_comparability/fieldwork_period_summary.csv`
- `q121_country_comparability/source_metadata.json`

## A2. Country and item comparability

Fieldwork began in 2024 for 16 samples and 2025 for eight. The country-level correlation between fieldwork midpoint and favorable evaluation is -.082; timing cannot be separated from country with this cross-section. The source exposes `I_PSU` for all 24 samples but no harmonized stratum field.

Of 192 thematic components, 170 have at least 20 valid observations in every country. Country-axis score coverage ranges from 82.1% to 100.0% under the 50% component threshold. The profile representation is stricter at the source level: a source must have at least 20 substantive observations in every included country.

Rerunning the profile models with only the 170 all-country components yields coefficient correlation .994, 95.9% sign agreement, and median absolute difference .009 relative to the primary scores. Restricting to the 16 samples whose fieldwork began in 2024 yields correlation .935, 86.8% sign agreement, and median difference .039. Science optimism remains omnibus-significant in both.

For country contexts, the harmonized-component coefficients correlate .990 with 92.4% sign agreement; science optimism and personal choice remain heterogeneous. The 2024-start coefficients correlate .873 with 87.9% sign agreement, and only science optimism remains omnibus-significant. These checks do not establish language invariance or isolate timing from country composition.

Artifacts:

- `q121_country_comparability/country_axis_component_coverage.csv`
- `q121_country_comparability/axis_component_country_observability.csv`
- `axis_comparability_sensitivity/coefficient_stability_summary.csv`
- `axis_comparability_sensitivity/README.md`

## A3. Predictor and thematic construction

The item-level prediction taxonomy contains ten literature-grounded classes. Eleven paper-facing thematic measures contain 192 retained indicators: one Q118 science item plus 191 indicators distributed across legitimacy, structural advantage, information, personal choice, agency, order, intermediary trust, redistribution, openness, and secular reasoning. They are formative summaries, not reflective latent scales.

All Q79 regional-organization items, including Q79_CAN, are excluded from prediction and clustering because their institutional referents differ across regions. Q240CS and Q241CS country-specific parental-education fields are excluded. Q147-Q154 use the source-native four-point agreement scale.

Q228 retains the source label “Respondent’s sex,” but the questionnaire asks which description most closely matches the respondent’s gender and notes that some samples use an older scale. Source codes are 1 = woman, 2 = man, 3 = transgender, 4 = non-binary, 5 = other, and 6 = prefer not to say; negative codes denote nonresponse or unavailability. The mapped sample contains 21,098, 19,571, 47, 53, 60, and 109 responses in these respective categories, with 22 missing. These are survey response categories, not evidence of sex assigned at birth, and national category availability is not harmonized. Prediction models one-hot encode categories 1-5 using training-fold encoding, treat code 6 as missing, and do not construct a country-relative numeric copy. Rare categories follow the same training-fold frequency rule as other nominal predictors. Missingness on Q228 does not exclude a respondent from outcome evaluation. Its SHAP contribution has no ordered gender interpretation.

In prediction, nondisclosure joins other Q228 nonresponse in the standard nominal missing-value category rather than receiving a position on a numeric scale. The separate outcome-independent profile representation already one-hot encoded Q228, with no country-relative copy. That frozen representation retains nondisclosure as a nominal source category. Profile assignments and thematic association models are held fixed in the prediction-coding sensitivity; Q228 is not a component of the thematic measures.

Within-context scores are constructed from fixed absolute item coding, centered on the WVS-weighted country mean, and divided by one equal-country pooled within-country SD. This single score is the main estimand. Separate absolute and country-relative coefficients are not interpreted as independent changes.

## A4. Prediction models

### A4.1 Pooled model and ablations

| Specification | Equal-country AUC | Change from full |
|---|---|---|
| Full clean model | .7689 | - |
| Absolute fields only | .7598 | -.0091 |
| Exclude Q118 | .7472 | -.0218 |
| Exclude science/technology family | .7462 | -.0227 |

The Q228 sensitivity uses the same 33,803 evaluators and the same three respondent splits. Mean equal-country AUC is .7685 with legacy numeric coding, .7689 with nominal coding, and .7682 with Q228 excluded. The legacy rerun exactly reproduces the archived split-level AUCs. Nominal coding is selected on measurement grounds, not its small AUC advantage; these matched predictive comparisons are not independent significance tests. In the seed-42 source-level SHAP comparison, the top-ten set is unchanged and Q228 moves from fourth to seventh. No directional gender claim is inferred from its nominal contribution. Aggregate results and the reproducible baseline configuration are in `q228_coding_sensitivity/`.

The duplicate-predictor sensitivity removes 35 derived recodes and indices, leaving 252 source predictors and a mean of 607 transformed columns. Its full AUC is .7675 (-.0014 relative to the primary model); its Q118 and science-family ablation losses are -.0201 and -.0212.

Primary artifacts:

- `q121_publication_models/pooled_ablation_summary.csv`
- `q121_publication_models/pooled_source_shap.csv`
- `q121_publication_models/figure1_clean_pooled_shap_beeswarm.png`
- `q121_publication_models_deduplicated/comparison_to_primary.md`

### A4.2 Separately fitted national models

Ten stratified respondent splits are fitted per country using absolute fields. Mean country AUC is .731, median AUC .731, and the range is .535-.836. Median pairwise top-10 Jaccard overlap is .271. Q118 enters at least one split's top ten in all 24 countries and at least half the splits in 20. Nominal Q228 reaches the half-of-splits threshold in five countries, followed by birth year and immigration evaluation in four each. No ordered Q228 direction is reported.

- `q121_country_source_recurrence/country_source_recurrence.csv`
- `q121_country_source_recurrence/source_recurrence_across_countries.csv`
- `q121_country_source_recurrence/figure_country_source_recurrence.png`
- `q121_publication_models/figure2_country_specific_shap_heatmap.png` (supplementary one-shot rank display)

### A4.3 Leave-one-country-out portability

LOCO uses only absolute fields because moments for a truly omitted country are unavailable during fitting. Macro country AUC is .654, median .673, and the range is .482-.747. Bangladesh, Egypt, Indonesia, and Morocco are below .55. LOCO SHAP describes transported predictions, not feature importance learned from the omitted sample.

- `q121_publication_models/loco_country_metrics.csv`
- `q121_publication_models/figure2_loco_shap_heatmap.png`

## A5. Corrected respondent representation and K selection

The corrected profile representation uses 243 source variables and 1,280 engineered features. Explicit exclusions remove 25 country, region, Q79, origin/language/ethnicity, and country-specific fields. A further 49 candidates fail the all-country observation rule. Missingness is neutralized in the geometry. Thirty-eight eligible sources receive country-relative copies; absolute and relative blocks each receive scale `1/sqrt(2)`.

| K | Mean graph ARI | Median Ward jump | Median silhouette | Median Davies-Bouldin | Minimum share in reference cut | Maximum share |
|---|---|---|---|---|---|---|
| 10 | .639 | 1.022 | .230 | 1.283 | 4.9% | 18.4% |
| 11 | .659 | 1.122 | .232 | 1.280 | 4.9% | 18.4% |
| 12 | .655 | 1.074 | .228 | 1.256 | 2.4% | 18.4% |
| 13 | .660 | 1.111 | .218 | 1.232 | 2.4% | 18.4% |
| 14 | .603 | 1.014 | .207 | 1.264 | 2.4% | 18.4% |
| 15 | .605 | 1.068 | .201 | 1.312 | 2.4% | 18.4% |

K=11 provides the best silhouette, the strongest merge discontinuity among the leading balanced cuts, and competitive graph stability. No AI outcome enters these diagnostics. Corrected alternative K=11 solutions have ARI .430-.678 with the primary and no cluster below 2%; the legacy partition has ARI .244.

Two additional constructions sample 1,000 respondents per country before preprocessing and embedding. Their ARI with primary labels for the same respondents is .425-.431; cluster shares range from 3.3% to 21.9%. Favorable evaluation spans 50.5%-67.6%, and between-profile evaluation variance remains .81%-1.01%.

- `ai_agnostic_proxyfree_natural_10d/seed42_proxyfree_pacmap10_variables_used.csv`
- `ai_agnostic_proxyfree_natural_10d/seed42_proxyfree_pacmap10_excluded_feature_catalog.csv`
- `ai_agnostic_proxyfree_natural_10d/seed42_proxyfree_pacmap10_variable_audit.csv`
- `ai_agnostic_proxyfree_natural_10d/natural_k_scan/natural_k_review.md`
- `ai_agnostic_proxyfree_natural_10d/partition_ensemble/`
- `ai_agnostic_proxyfree_equal_country/equal_country_partition_comparison.csv`
- `ai_agnostic_proxyfree_equal_country/README.md`

## A6. Respondent profiles and variance decomposition

| Profile | N | Share | WVS favorable | Largest country share |
|---|---|---|---|---|
| C0 Secular cosmopolitan choice liberals | 4,062 | 9.9% | 54.1% | 33.5% CAN |
| C1 Secular private-choice individualists | 3,803 | 9.3% | 62.2% | 19.9% CAN |
| C2 Religious social traditionalists | 3,618 | 8.8% | 49.1% | 12.1% ZAF |
| C3 Secular moderate pluralists | 3,614 | 8.8% | 55.2% | 13.7% CAN |
| C4 Disadvantaged low-participation order seekers | 3,413 | 8.3% | 53.1% | 13.8% ZAF |
| C5 Faith-centered institutional mainstream | 5,198 | 12.7% | 60.6% | 14.7% ZAF |
| C6 Religious restrictive material solidarists | 7,550 | 18.4% | 60.3% | 16.3% IDN |
| C7 Transgression-tolerant coercive-order outliers | 2,071 | 5.1% | 59.0% | 15.9% ZAF |
| C8 Boundary-defensive localists | 2,014 | 4.9% | 51.1% | 14.1% CAN |
| C9 Institution-trusting civic joiners | 3,155 | 7.7% | 56.4% | 25.4% ZAF |
| C10 Agentic associational traditionalists | 2,462 | 6.0% | 60.7% | 12.1% ZAF |

Between-profile means account for .70% of WVS-weighted individual evaluation variance and .69% under equal-country weighting.

- `proxyfree_profile_summaries/profile_publication_summary.csv`
- `proxyfree_profile_summaries/profile_variance_decomposition.csv`
- `proxyfree_profile_summaries/profile_summaries.md`

## A7. Profile interaction models

The formal model includes 33,045 evaluators, 11 profiles, 11 jointly entered thematic scores, country fixed intercepts, and equal-country WVS weights. There are 24 country-delete refits and no limited-support cell among 121 profile-axis combinations.

Primary omnibus heterogeneity survives BH correction for eight axes. Across three corrected alternative partitions, recurrence is:

| Axis | Significant alternatives (of 3) |
|---|---|
| Science/technology optimism | 3 |
| Redistributive protection | 2 |
| Secular reasoning | 2 |
| Institutional legitimacy | 1 |

| Axis | Significant alternatives (of 3) |
|---|---|
| Structural advantage | 1 |
| Information connectedness | 1 |
| Coercive order | 1 |
| Intermediary trust | 1 |
| Personal-choice liberalism | 0 |
| Human agency | 0 |
| Boundary openness | 0 |

Alternative-partition coefficient correlations are .852-.954 and sign agreement is 83.5%-90.9%. The cross-classified ridge sensitivity has coefficient correlation .950 and sign agreement 92.6% with the primary model. These are stability diagnostics, not post-selection-adjusted inference.

- `profile_axis_ai_within_country_publication/profile_axis_ai_slopes.csv`
- `profile_axis_ai_within_country_publication/profile_axis_ai_omnibus_tests.csv`
- `profile_axis_ai_within_country_publication/alternative_partition_omnibus_tests.csv`
- `profile_axis_ai_within_country_publication/profile_axis_support_diagnostics.csv`
- `profile_axis_ai_within_country_publication/figure_appendix_profile_axis_intervals.png`
- `profile_axis_ai_within_country_publication/figure_appendix_profile_axis_probability_effects.png`
- `profile_axis_ai_within_country_publication/figure_appendix_profile_axis_support.png`
- `profile_axis_ai_within_country_publication/figure_appendix_partition_sensitivity.png`
- `profile_axis_ai_within_country_publication/figure_appendix_cross_classified_sensitivity.png`

## A8. PaCMAP overlap display

The two-dimensional display is refitted from the same corrected proxy-free source set with PaCMAP `n_neighbors=50`. Q121 is attached only after coordinates are frozen. Favorable and unfavorable weighted histograms are convolved using one pooled Scott bandwidth (0.50 by 0.56 PaCMAP units). The local favorable ratio is masked outside the highest-density region containing 99% of evaluator density. Integrating the smoothed ratio against pooled density yields 57.37%, close to the observed WVS-weighted 57.36%.

- `pacmap2d_local_ai_support_frequency/figure3b_pacmap_support_oppose_kde_ratio.png`
- `pacmap2d_local_ai_support_frequency/pacmap_kde_ratio_grid.csv` (aggregate grid in the research materials)
- `pacmap2d_local_ai_support_frequency/kde_run_metadata.json`

## A9. Country contexts

The primary K=3 hierarchy contains no singleton. Group sizes are 4, 10, and 10 countries; silhouette is .300. Mean respondent-bootstrap ARI is .975, PAM ARI .780, profile-only ARI 1.000, axis-only ARI .733, and Aitchison-versus-Hellinger ARI .808. Changing the profile/axis weights to 35/65 or 65/35 leaves assignments unchanged.

| Context | Countries | Favorable among evaluators |
|---|---|---|
| R1 Religious-traditional | BGD, EGY, IDN, MAR | 65.7% |
| R2 Mixed traditional-organizational | BOL, CIV, COL, GTM, KEN, MYS, NGA, PAK, PER, ZAF | 57.3% |
| R3 Secular-emancipative | AND, CAN, GRC, JPN, SGP, SVK, SWE, THA, TWN, USA | 61.0% |

No axis has corrected omnibus context heterogeneity for narrow evaluability. Science optimism (`q<.001`) and personal-choice liberalism (`q=.032`) differ across contexts for favorable evaluation. Secular reasoning is borderline (`q=.052`); other omnibus tests are not significant.

- `country_value_context_clustering/review.md`
- `country_value_context_clustering/k_diagnostics.csv`
- `country_context_axis_ai_within_country_publication/country_context_axis_slopes.csv`
- `country_context_axis_ai_within_country_publication/country_context_axis_omnibus_tests.csv`
- `country_context_axis_ai_within_country_publication/figure_appendix_country_context_axis_intervals.png`
- `country_context_axis_ai_within_country_publication/figure_appendix_country_context_axis_probability_effects.png`
- `country_context_axis_ai_within_country_publication/figure_appendix_country_context_axis_support.png`

## A10. Decision history and evidentiary status

The analysis was iterative and used substantial agentic AI assistance. OpenAI Codex and ChatGPT supported methodological ideation, development and debugging of analysis and visualization code, and manuscript drafting and revision. Earlier pilots included Q121-weighted clustering, HDBSCAN variants, outcome-regularized merging, subgroup discovery, and GLM-5.2 (`z-ai/glm-5.2`) API-assisted interpretation of clusters, predictors, and thematic associations. Those pilots informed method development and interpretation but do not provide independent numerical or inferential evidence for the reported results. The publication design fixes these rules:

1. Q121 does not enter respondent or country clustering.
2. Country and regional proxies, all Q79 items, structurally country-specific fields, and source-availability signals are excluded from profile geometry.
3. Missingness does not contribute to distance.
4. K is assessed with outcome-free diagnostics and a prespecified 10-15 range.
5. Profile association claims are checked across three corrected partitions.
6. Country contexts use profile shares and mean thematic positions only.
7. Language-model prose is not a substitute for empirical estimates or statistical tests.

The resulting interaction tests remain exploratory and conditional on learned partitions. BH correction addresses stated test families but not all researcher degrees of freedom accumulated during development.

## A11. Reproducibility and research materials

Public version 1.0 of the research materials is archived in Zenodo: Flanders, S., and M. Nungsari (2026), *Research Materials for Conditional Values: The Contingent Foundations of Attitudes toward Artificial Intelligence*, https://doi.org/10.5281/zenodo.23688866. These materials contain deterministic analysis scripts, operational metadata, fixed annotations and item maps, aggregate output tables, software requirements, and a SHA-256 integrity manifest. Original code is MIT-licensed and original documentation and outputs are CC BY 4.0; attributed WVS and other third-party content are excluded from those grants. Raw surveys, respondent assignments, individual predictions, and individual SHAP records are not redistributed. Obtain the exact provisional SAV independently under WVS access conditions and place it at `raw files/WVS8_v1_Spss_eng_v20260421.sav`.

Prepare the Wave 8 input without requiring earlier-wave data:

```
python .\scripts\prepare_ai_submission_source.py
```

The supplied column dictionary maps 353 native source fields and one derived education indicator (Q239 at bachelor level or higher). All 354 mapped fields and all 40,960 rows were checked against the publication analysis input with exact agreement. The preserved index offset aligns regenerated records with the existing pipeline; it does not enter prediction or clustering.

The deterministic regeneration entry point is:

```
python .\scripts\run_ai_paper_publication_revision.py
```

The driver can skip existing prediction or embedding stages with `--skip-prediction` or `--skip-embeddings`. A fresh replication must run both stages. No generative-AI API calls are required. Analysis regeneration leaves the editorial manuscript unchanged unless the template-overwrite option is explicitly requested. The supplement README distinguishes fresh statistical replication from re-exporting the submission documents.

## A12. Declarations

**Ethics.** This study uses secondary, de-identified World Values Survey data and involved no new participant recruitment or data collection. Institutional review board review is not required under the authors' institutional policy for secondary analysis of existing sources such as the WVS.

**Funding.** This research did not receive any specific grant from funding agencies in the public, commercial, or not-for-profit sectors.

**Competing interests.** The authors declare no competing interests.

Preprint status. This author manuscript has not been peer reviewed.

**AI assistance.** The research used substantial agentic AI support for coding, methodological ideation, exploratory interpretation, and manuscript preparation. Tool use is described in manuscript Section 3.8 and the separate AI declaration; it is not limited to language editing.

Author contributions. Sam Flanders: Conceptualization; Methodology; Software; Validation; Formal analysis; Investigation; Resources; Data curation; Writing - original draft; Writing - review and editing; Visualization; Supervision; Project administration.

Melati Nungsari: Conceptualization; Methodology; Writing - review and editing.

Code and aggregate outputs are publicly archived in Zenodo at https://doi.org/10.5281/zenodo.23688866, with the source-data access restriction and license boundaries described in A11.

## A13. Extended association displays

Figures S1A-S1B expand the compact main coefficient display into profile-specific intervals. Figures S2A-S2B express the same joint-model associations as probability differences. Figures S3A-S3C report observed range and effective component variation, distinguishing information available within a profile from the estimated association. Figures S4A-S4B provide intervals for favorable evaluation in the country contexts. All displays use frozen publication tables; no models or clustering were refitted for these exports.

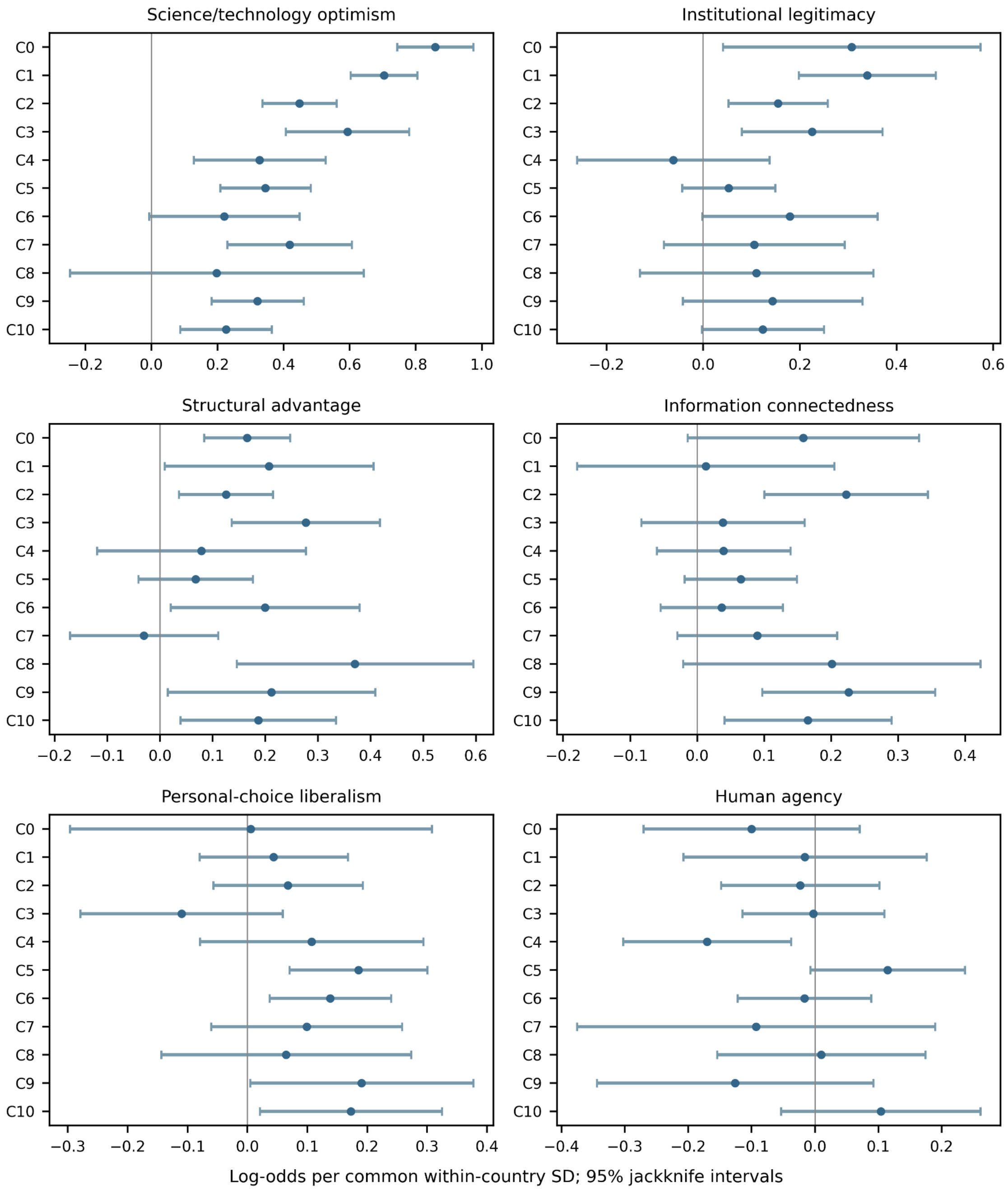

*Figure S1A. Profile-specific log-odds coefficients for the first six thematic measures. Intervals are 95% country-delete jackknife intervals using a t reference with 23 degrees of freedom. Every coefficient is per one common pooled within-country standard deviation (SD), jointly adjusting for the other measures and country intercepts. These intervals are not simultaneous confidence intervals; corrected significance is reported in the main display and machine-readable tables. Plotting code was developed with AI assistance.*

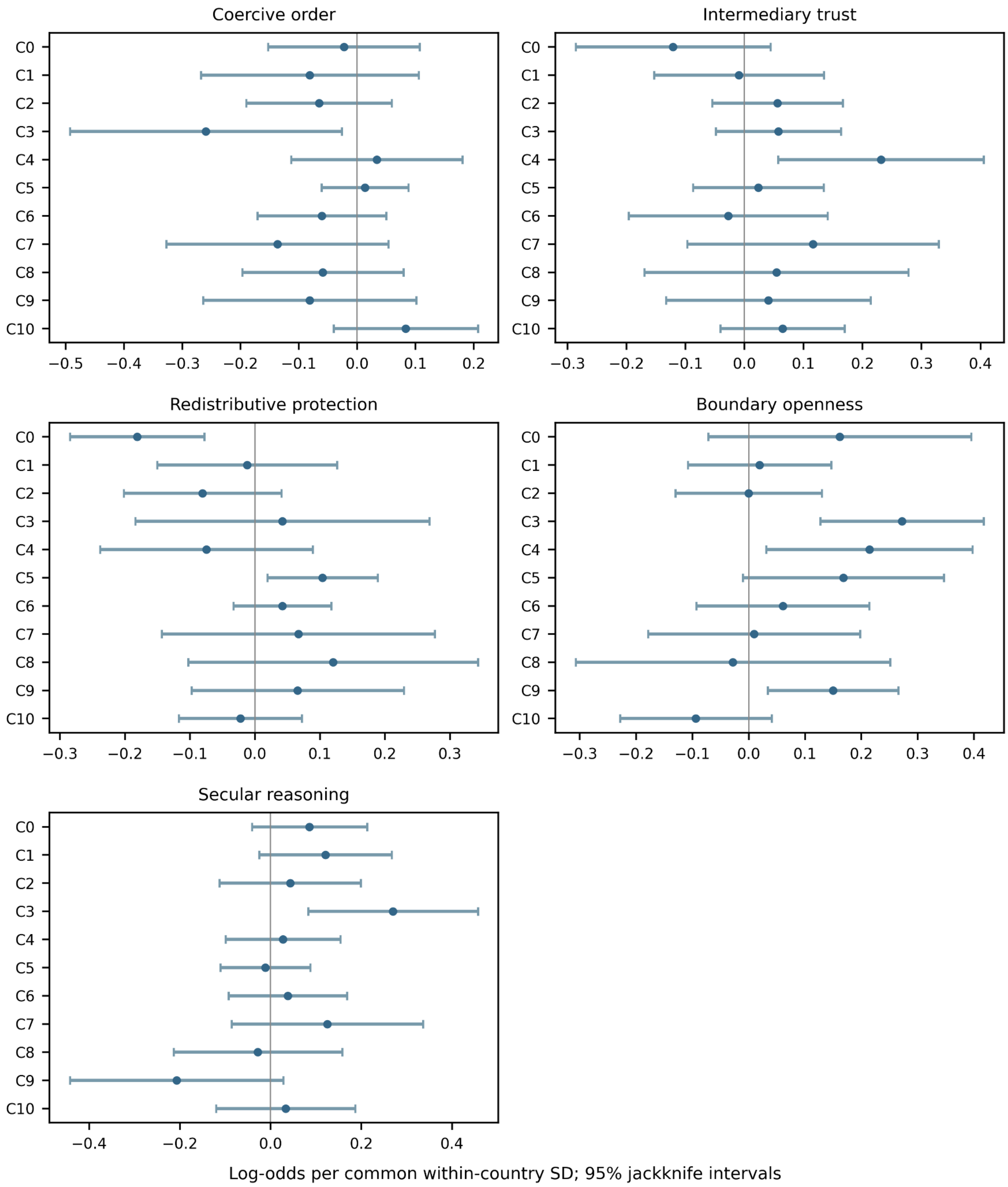

*Figure S1B. Profile-specific log-odds coefficients for the remaining five thematic measures, with the same common-scale definition and uncertainty calculation as Figure S1A. Plotting code was developed with AI assistance.*

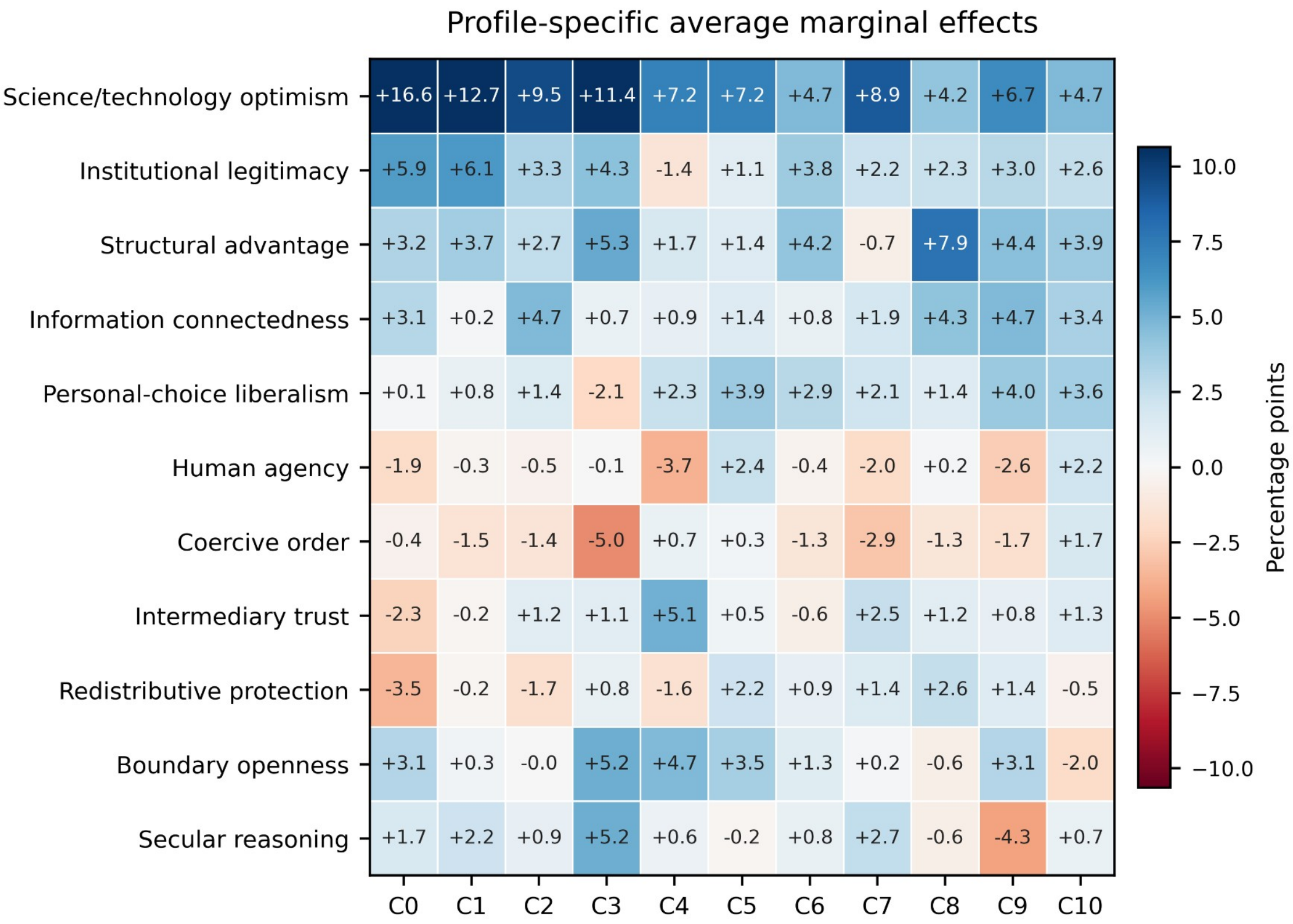

*Figure S2A. Profile-specific average marginal effects, in percentage points per common within-country SD. These are adjusted associations, not causal effects. The joint-model log-odds tests remain the primary inferential analysis. Plotting code was developed with AI assistance.*

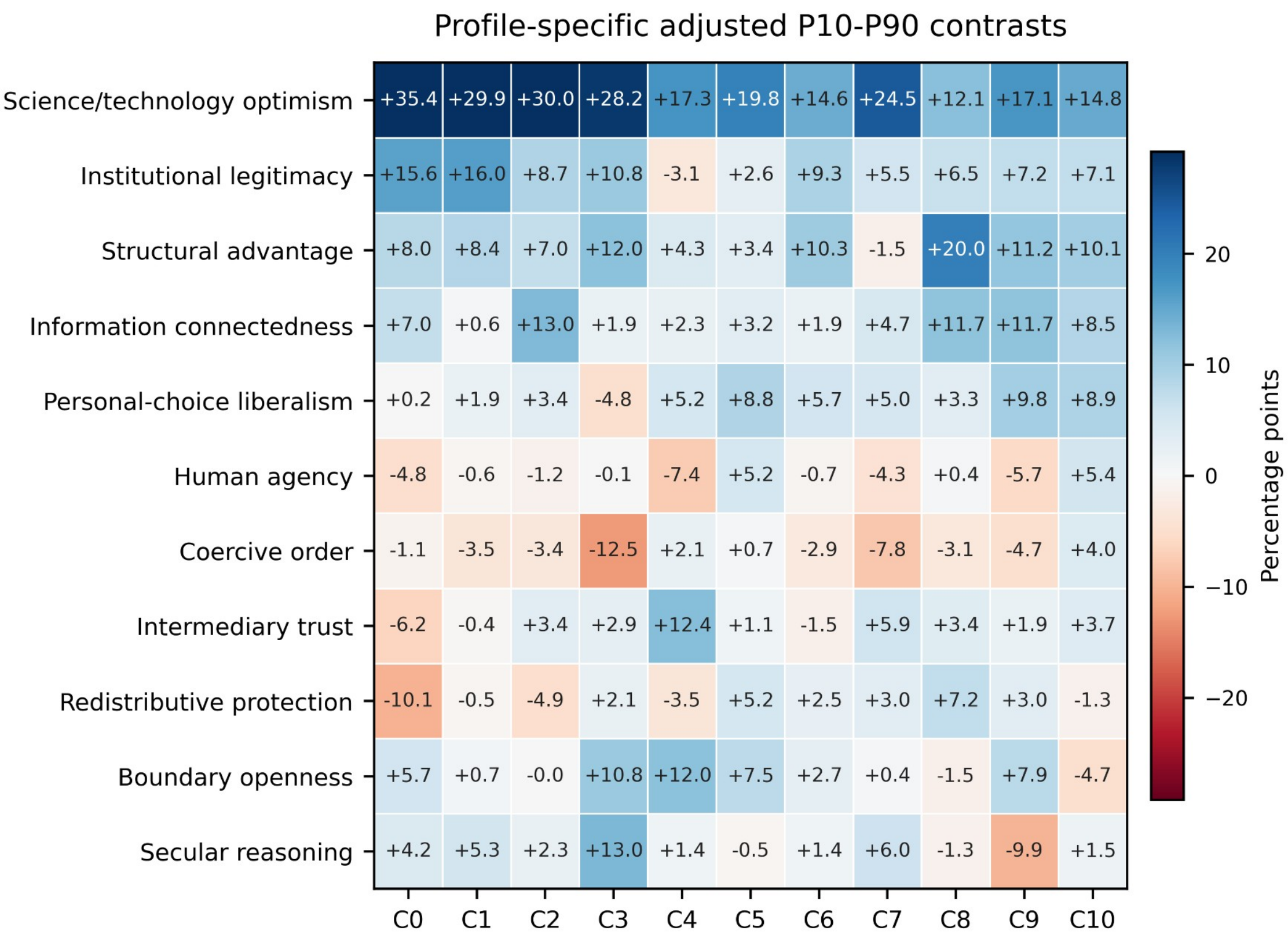

*Figure S2B. Adjusted favorable-evaluation probability differences between each profile's observed 10th and 90th percentile of a thematic score (P10-P90). Contrasts reflect both the estimated slope and the profile's observed range; they do not represent equal-sized changes across profiles. Plotting code was developed with AI assistance.*

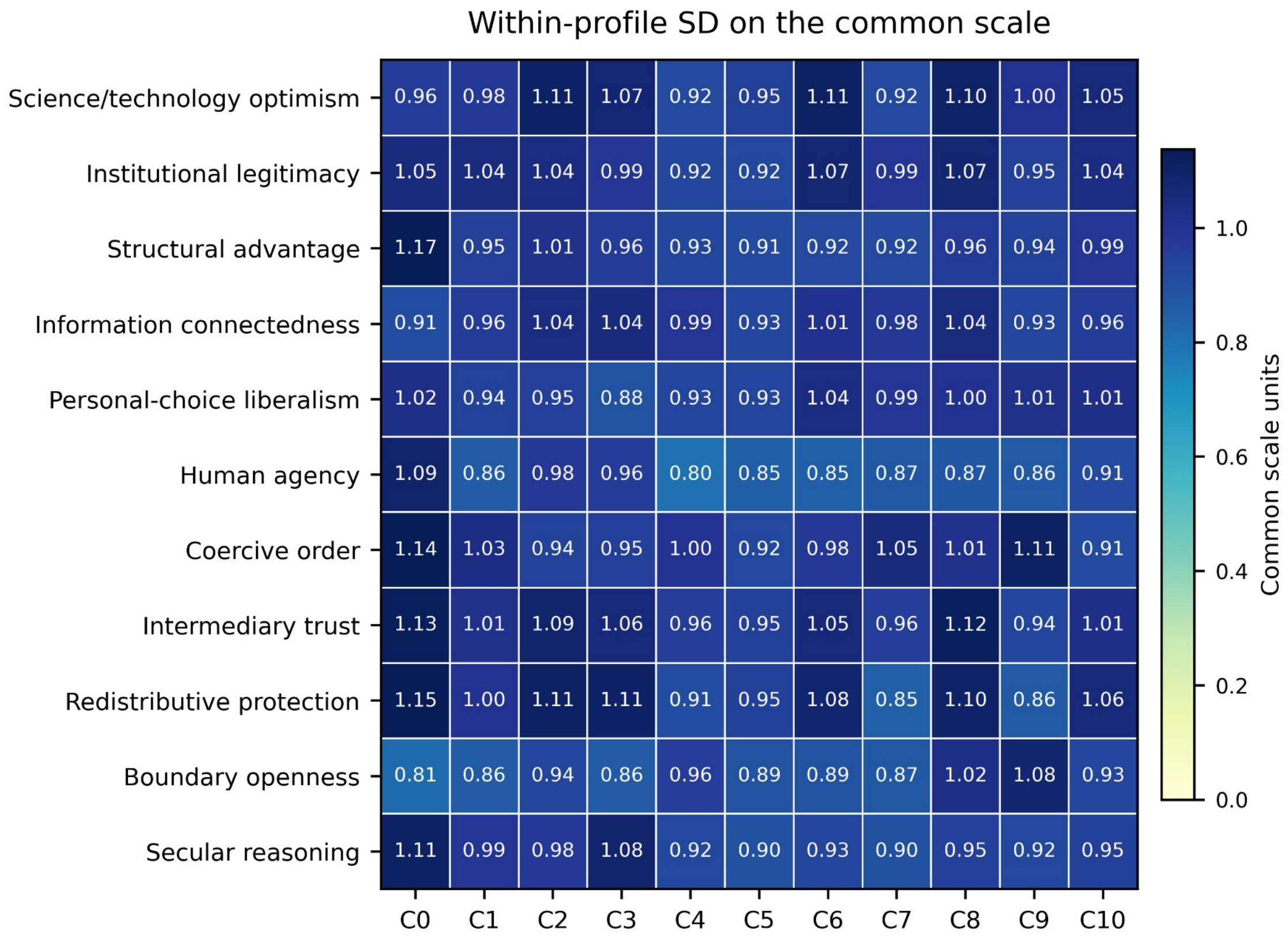

*Figure S3A. Within-profile score SD measured on the common pooled within-country scale. Values describe the variation available to estimate the association; they do not indicate coefficient strength. Plotting code was developed with AI assistance.*

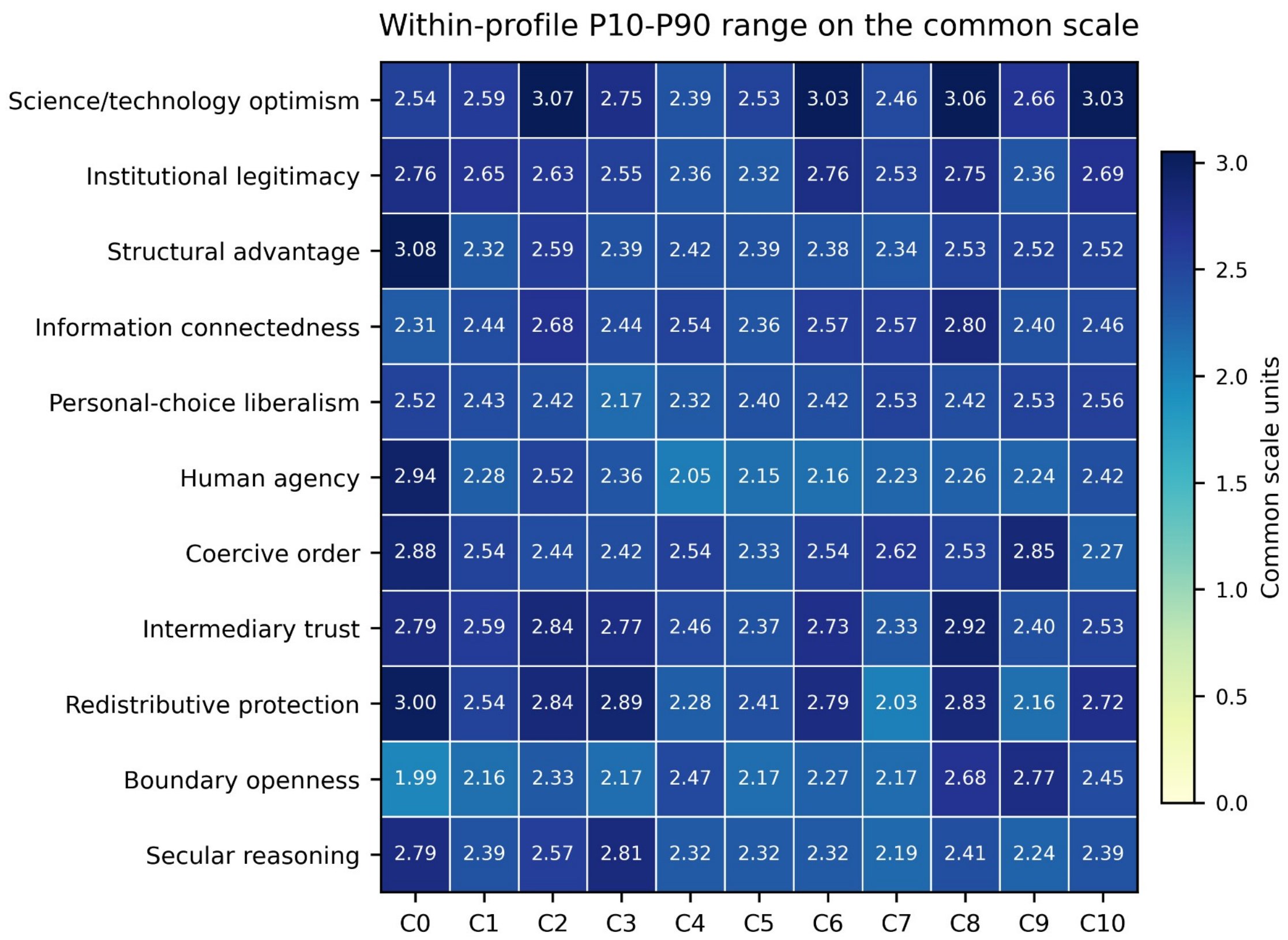

*Figure S3B. Within-profile P10-P90 score range on the same common scale. Narrow ranges would limit interpretation of an association; sparse-tail diagnostics are provided in the research materials. Plotting code was developed with AI assistance.*

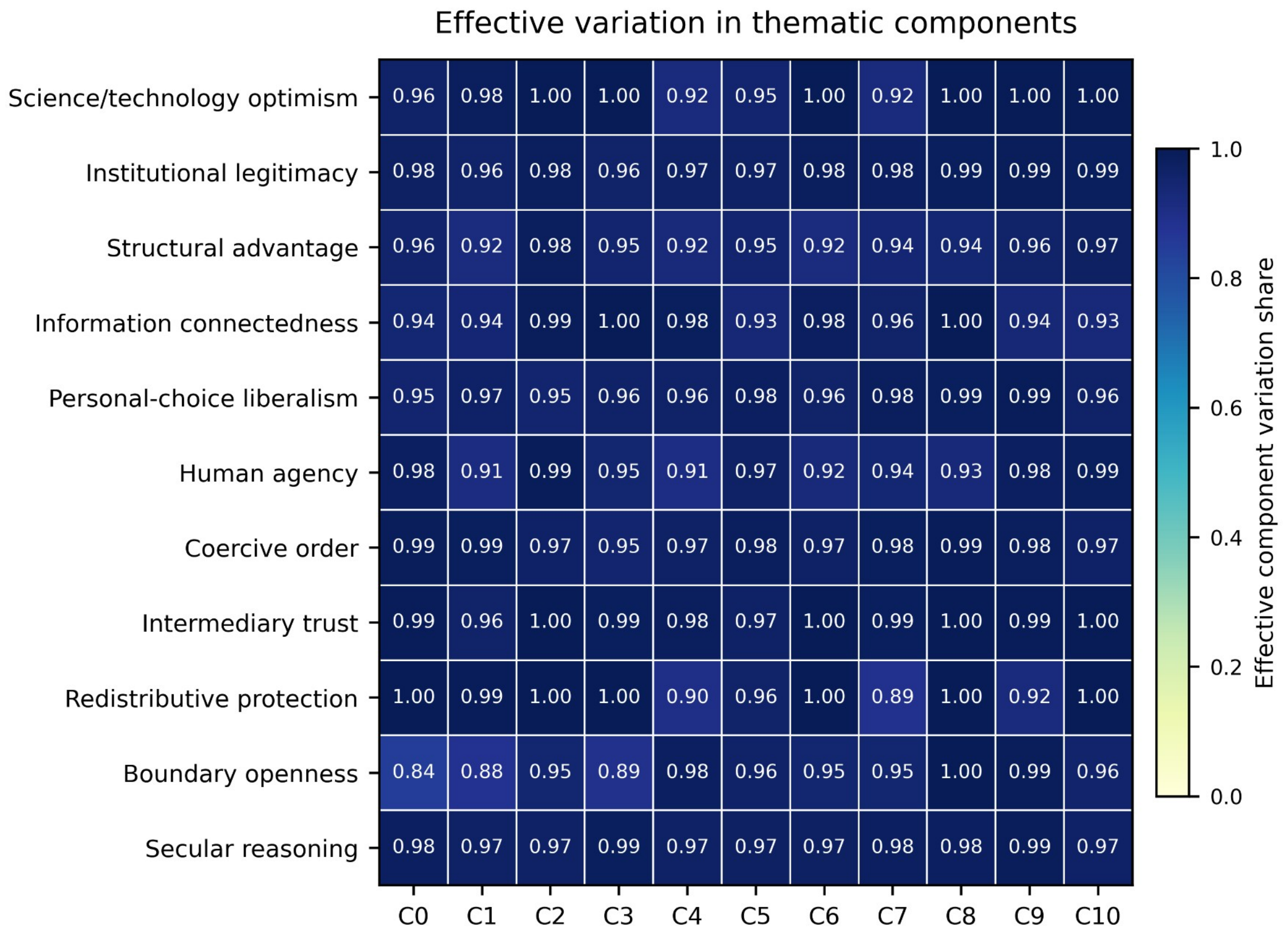

*Figure S3C. Effective component variation: the mean within-profile component SD on the common scale, with each component capped at one before averaging. This is a variation diagnostic, not the fraction of components that vary. The single-item science measure is included. No primary profile-measure cell meets the prespecified limited-support flag. The complete component-level diagnostics accompany the research materials. Plotting code was developed with AI assistance.*

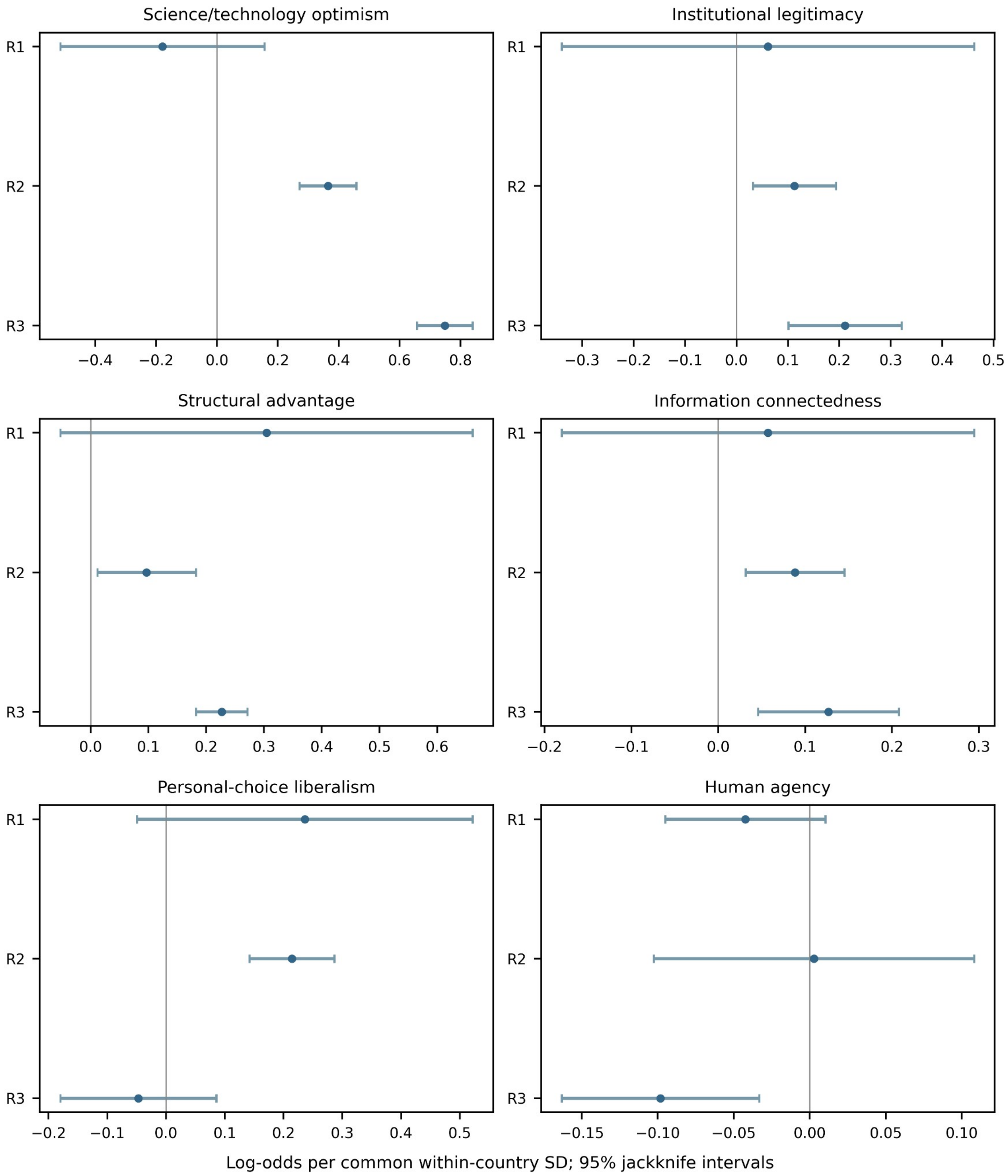

*Figure S4A. Joint favorable-evaluation coefficients for the first six measures within the country contexts R1-R3. The common within-country score unit, country intercept adjustment, and 95% country-delete jackknife intervals match the manuscript analysis. The outcome is favorable versus unfavorable evaluation, not narrow evaluability. Plotting code was developed with AI assistance.*

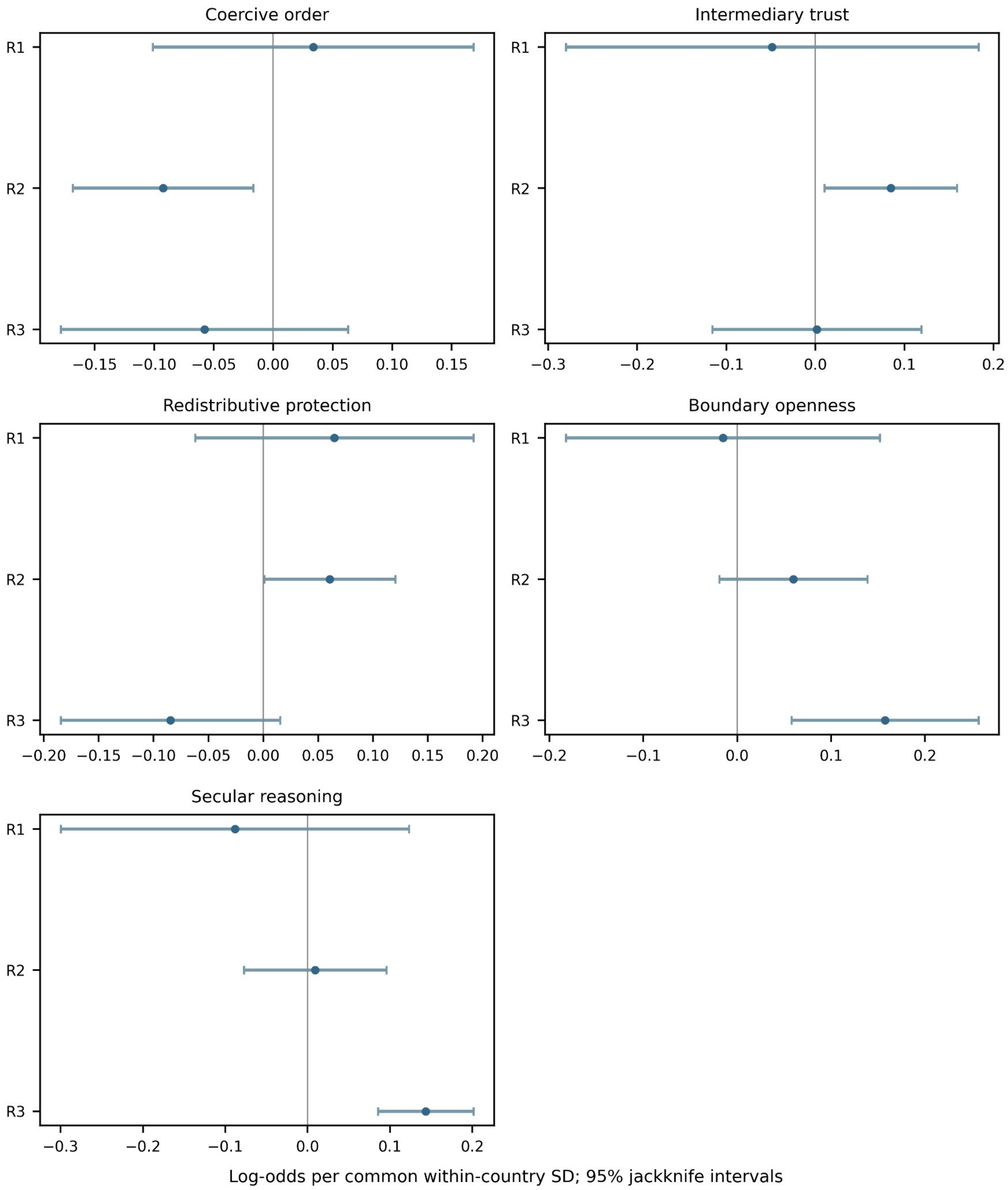

*Figure S4B. Joint favorable-evaluation coefficients for the remaining five measures, with the same definition and uncertainty calculation as Figure S4A. Narrow-evaluability estimates, probability contrasts, and diagnostics are supplied as machine-readable tables in the research materials. Plotting code was developed with AI assistance.*